\documentclass[preprint
  ,prd
  ,tightenlines
  ,superscriptaddress]{revtex4-1}

\usepackage{graphicx} 
\usepackage{dcolumn}  
\usepackage{colordvi}
\usepackage{color}
\usepackage{pstricks}
\usepackage{epstopdf}
\usepackage{amssymb}
\usepackage{url}
\graphicspath{{ps}}
\usepackage{hyperref}
\usepackage{tabularx}
\usepackage{multirow}
\usepackage{units}
\usepackage{siunitx}
\usepackage{hyphenat}
\usepackage{listings}
\usepackage{subfigure}
\usepackage[italic]{hepnames}

\renewcommand{\PBzero}{\ensuremath{\HepParticle{\PB}{}{}^0}\xspace}
\renewcommand{\APBzero}{\ensuremath{\HepParticle{\APB}{}{}^0}\xspace}

\renewcommand{\APDzero}{\ensuremath{\HepParticle{\APD}{}{}^0}\xspace}

\renewcommand{\PKzS}{\ensuremath{\HepParticle{\PK}{}{}^0_{\rm S}}\xspace}

\input ./belle2sym.tex

\begin{document}

\def\belletwo {\it {Belle II}}

\clubpenalty = 10000  
\widowpenalty = 10000 

\vspace*{-3\baselineskip}
\resizebox{!}{3cm}{\includegraphics{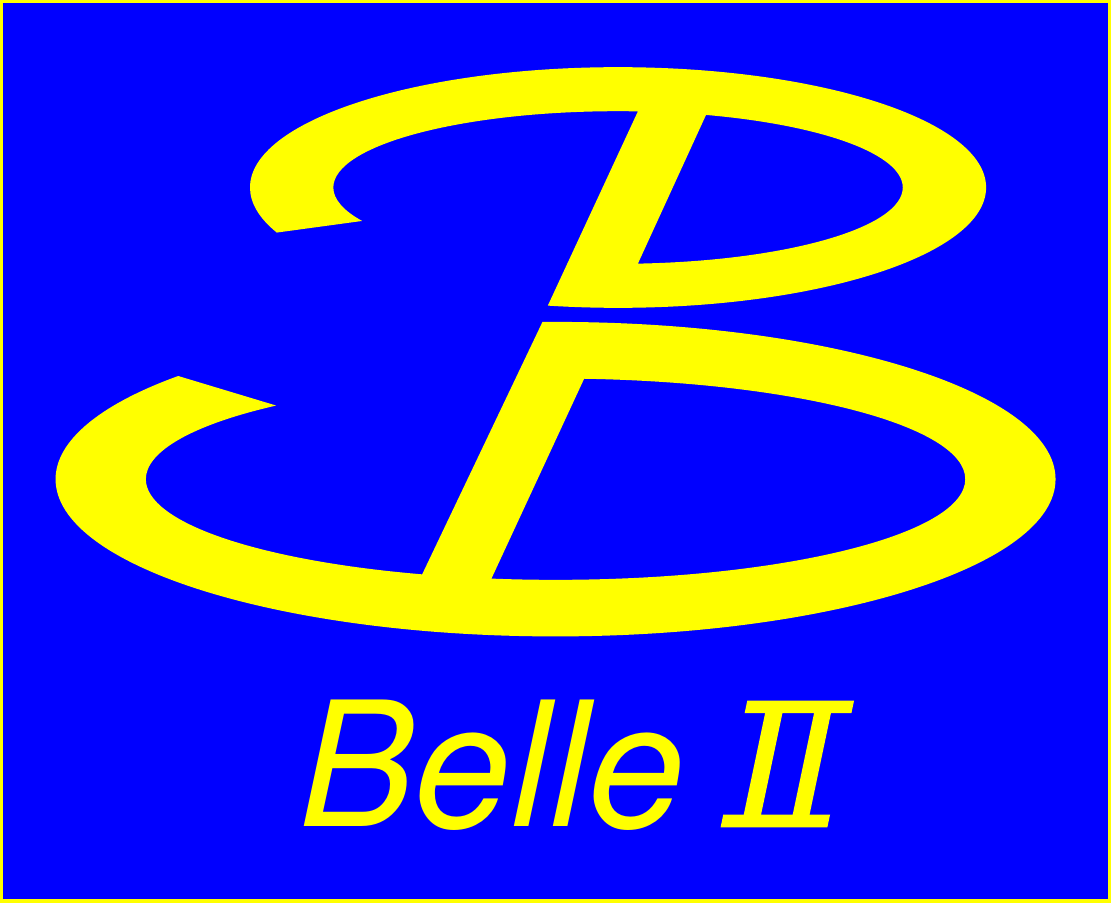}}

\vspace*{-5\baselineskip}
\begin{flushright}
BELLE2-NOTE-PH-2021-010\\
\today
\end{flushright}

\title { \quad\\[0.5cm] Measurements of branching fractions and direct CP asymmetries in $B^{0}\to K^{+} \pi^{-}$, $B^+ \to \PKzS\pi^+$ and $B^0 \to \pi^+\pi^-$ using 2019 and 2020 data}
\newcommand{\instCPPM}{Aix Marseille Universit\'{e}, CNRS/IN2P3, CPPM, 13288 Marseille, France}
\newcommand{\instBeihang}{Beihang University, Beijing 100191, China}
\newcommand{\instBNL}{Brookhaven National Laboratory, Upton, New York 11973, U.S.A.}
\newcommand{\instBINP}{Budker Institute of Nuclear Physics SB RAS, Novosibirsk 630090, Russian Federation}
\newcommand{\instCMU}{Carnegie Mellon University, Pittsburgh, Pennsylvania 15213, U.S.A.}
\newcommand{\instCinvestavIPN}{Centro de Investigacion y de Estudios Avanzados del Instituto Politecnico Nacional, Mexico City 07360, Mexico}
\newcommand{\instPrague}{Faculty of Mathematics and Physics, Charles University, 121 16 Prague, Czech Republic}
\newcommand{\instChiangMai}{Chiang Mai University, Chiang Mai 50202, Thailand}
\newcommand{\instChiba}{Chiba University, Chiba 263-8522, Japan}
\newcommand{\instChonnam}{Chonnam National University, Gwangju 61186, South Korea}
\newcommand{\instConacyt}{Consejo Nacional de Ciencia y Tecnolog\'{\i}a, Mexico City 03940, Mexico}
\newcommand{\instDESY}{Deutsches Elektronen--Synchrotron, 22607 Hamburg, Germany}
\newcommand{\instDuke}{Duke University, Durham, North Carolina 27708, U.S.A.}
\newcommand{\instITAR}{ Duy Tan University, Hanoi 100000, Vietnam}
\newcommand{\instRomaENEA}{ENEA Casaccia, I-00123 Roma, Italy}
\newcommand{\instEri}{Earthquake Research Institute, University of Tokyo, Tokyo 113-0032, Japan}
\newcommand{\instJuelich}{Forschungszentrum J\"{u}lich, 52425 J\"{u}lich, Germany}
\newcommand{\instFuJen}{Department of Physics, Fu Jen Catholic University, Taipei 24205, Taiwan}
\newcommand{\instFudan}{Key Laboratory of Nuclear Physics and Ion-beam Application (MOE) and Institute of Modern Physics, Fudan University, Shanghai 200443, China}
\newcommand{\instGoettingen}{II. Physikalisches Institut, Georg-August-Universit\"{a}t G\"{o}ttingen, 37073 G\"{o}ttingen, Germany}
\newcommand{\instGifu}{Gifu University, Gifu 501-1193, Japan}
\newcommand{\instSOKENDAI}{The Graduate University for Advanced Studies (SOKENDAI), Hayama 240-0193, Japan}
\newcommand{\instGyeongsang}{Gyeongsang National University, Jinju 52828, South Korea}
\newcommand{\instHanyang}{Department of Physics and Institute of Natural Sciences, Hanyang University, Seoul 04763, South Korea}
\newcommand{\instKEK}{High Energy Accelerator Research Organization (KEK), Tsukuba 305-0801, Japan}
\newcommand{\instJPARC}{J-PARC Branch, KEK Theory Center, High Energy Accelerator Research Organization (KEK), Tsukuba 305-0801, Japan}
\newcommand{\instHSE}{Higher School of Economics (HSE), Moscow 101000, Russian Federation}
\newcommand{\instIISER}{Indian Institute of Science Education and Research Mohali, SAS Nagar, 140306, India}
\newcommand{\instIITBhubaneswar}{Indian Institute of Technology Bhubaneswar, Satya Nagar 751007, India}
\newcommand{\instIITGuwahati}{Indian Institute of Technology Guwahati, Assam 781039, India}
\newcommand{\instIITHyderabad}{Indian Institute of Technology Hyderabad, Telangana 502285, India}
\newcommand{\instIITMadras}{Indian Institute of Technology Madras, Chennai 600036, India}
\newcommand{\instIndiana}{Indiana University, Bloomington, Indiana 47408, U.S.A.}
\newcommand{\instIHEPRussia}{Institute for High Energy Physics, Protvino 142281, Russian Federation}
\newcommand{\instHEPHYVienna}{Institute of High Energy Physics, Vienna 1050, Austria}
\newcommand{\instHiroshima}{Hiroshima University, Higashi-Hiroshima, Hiroshima 739-8530, Japan}
\newcommand{\instIHEPChina}{Institute of High Energy Physics, Chinese Academy of Sciences, Beijing 100049, China}
\newcommand{\instIPP}{Institute of Particle Physics (Canada), Victoria, British Columbia V8W 2Y2, Canada}
\newcommand{\instIOP}{Institute of Physics, Vietnam Academy of Science and Technology (VAST), Hanoi, Vietnam}
\newcommand{\instIFIC}{Instituto de Fisica Corpuscular, Paterna 46980, Spain}
\newcommand{\instFrascati}{INFN Laboratori Nazionali di Frascati, I-00044 Frascati, Italy}
\newcommand{\instNapoliINFN}{INFN Sezione di Napoli, I-80126 Napoli, Italy}
\newcommand{\instPadovaINFN}{INFN Sezione di Padova, I-35131 Padova, Italy}
\newcommand{\instPerugiaINFN}{INFN Sezione di Perugia, I-06123 Perugia, Italy}
\newcommand{\instPisaINFN}{INFN Sezione di Pisa, I-56127 Pisa, Italy}
\newcommand{\instRomaINFN}{INFN Sezione di Roma, I-00185 Roma, Italy}
\newcommand{\instRomaTreINFN}{INFN Sezione di Roma Tre, I-00146 Roma, Italy}
\newcommand{\instTorinoINFN}{INFN Sezione di Torino, I-10125 Torino, Italy}
\newcommand{\instTriesteINFN}{INFN Sezione di Trieste, I-34127 Trieste, Italy}
\newcommand{\instJAEA}{Advanced Science Research Center, Japan Atomic Energy Agency, Naka 319-1195, Japan}
\newcommand{\instMainz}{Johannes Gutenberg-Universit\"{a}t Mainz, Institut f\"{u}r Kernphysik, D-55099 Mainz, Germany}
\newcommand{\instGiessen}{Justus-Liebig-Universit\"{a}t Gie\ss{}en, 35392 Gie\ss{}en, Germany}
\newcommand{\instKarlsruhe}{Institut f\"{u}r Experimentelle Teilchenphysik, Karlsruher Institut f\"{u}r Technologie, 76131 Karlsruhe, Germany}
\newcommand{\instISU}{Iowa State University, Ames, Iowa 50011, U.S.A.}
\newcommand{\instKitasato}{Kitasato University, Sagamihara 252-0373, Japan}
\newcommand{\instKISTI}{Korea Institute of Science and Technology Information, Daejeon 34141, South Korea}
\newcommand{\instKoreaUnivKU}{Korea University, Seoul 02841, South Korea}
\newcommand{\instKSU}{Kyoto Sangyo University, Kyoto 603-8555, Japan}
\newcommand{\instKyungpook}{Kyungpook National University, Daegu 41566, South Korea}
\newcommand{\instLPI}{P.N. Lebedev Physical Institute of the Russian Academy of Sciences, Moscow 119991, Russian Federation}
\newcommand{\instLNNU}{Liaoning Normal University, Dalian 116029, China}
\newcommand{\instLMU}{Ludwig Maximilians University, 80539 Munich, Germany}
\newcommand{\instLuther}{Luther College, Decorah, Iowa 52101, U.S.A.}
\newcommand{\instMNITJaipur}{Malaviya National Institute of Technology Jaipur, Jaipur 302017, India}
\newcommand{\instMPP}{Max-Planck-Institut f\"{u}r Physik, 80805 M\"{u}nchen, Germany}
\newcommand{\instMPGHLL}{Semiconductor Laboratory of the Max Planck Society, 81739 M\"{u}nchen, Germany}
\newcommand{\instMcGill}{McGill University, Montr\'{e}al, Qu\'{e}bec, H3A 2T8, Canada}
\newcommand{\instMEPhI}{Moscow Physical Engineering Institute, Moscow 115409, Russian Federation}
\newcommand{\instNagoya}{Graduate School of Science, Nagoya University, Nagoya 464-8602, Japan}
\newcommand{\instNagoyaIAR}{Institute for Advanced Research, Nagoya University, Nagoya 464-8602, Japan}
\newcommand{\instNagoyaKMI}{Kobayashi-Maskawa Institute, Nagoya University, Nagoya 464-8602, Japan}
\newcommand{\instNaraWu}{Nara Women's University, Nara 630-8506, Japan}
\newcommand{\instNTUTaiwan}{Department of Physics, National Taiwan University, Taipei 10617, Taiwan}
\newcommand{\instNUUTaiwan}{National United University, Miao Li 36003, Taiwan}
\newcommand{\instKrakow}{H. Niewodniczanski Institute of Nuclear Physics, Krakow 31-342, Poland}
\newcommand{\instNiigata}{Niigata University, Niigata 950-2181, Japan}
\newcommand{\instNSU}{Novosibirsk State University, Novosibirsk 630090, Russian Federation}
\newcommand{\instOkinawa}{Okinawa Institute of Science and Technology, Okinawa 904-0495, Japan}
\newcommand{\instOsakaCity}{Osaka City University, Osaka 558-8585, Japan}
\newcommand{\instRCNP}{Research Center for Nuclear Physics, Osaka University, Osaka 567-0047, Japan}
\newcommand{\instPNNL}{Pacific Northwest National Laboratory, Richland, Washington 99352, U.S.A.}
\newcommand{\instPanjab}{Panjab University, Chandigarh 160014, India}
\newcommand{\instPanjabPAU}{Punjab Agricultural University, Ludhiana 141004, India}
\newcommand{\instRIKENMSL}{Meson Science Laboratory, Cluster for Pioneering Research, RIKEN, Saitama 351-0198, Japan}
\newcommand{\instSeoul}{Seoul National University, Seoul 08826, South Korea}
\newcommand{\instSPU}{Showa Pharmaceutical University, Tokyo 194-8543, Japan}
\newcommand{\instSoochow}{Soochow University, Suzhou 215006, China}
\newcommand{\instSoongsil}{Soongsil University, Seoul 06978, South Korea}
\newcommand{\instLjubljanaJSI}{J. Stefan Institute, 1000 Ljubljana, Slovenia}
\newcommand{\instKyiv}{Taras Shevchenko National Univ. of Kiev, Kiev, Ukraine}
\newcommand{\instTata}{Tata Institute of Fundamental Research, Mumbai 400005, India}
\newcommand{\instTUM}{Department of Physics, Technische Universit\"{a}t M\"{u}nchen, 85748 Garching, Germany}
\newcommand{\instTelAviv}{Tel Aviv University, School of Physics and Astronomy, Tel Aviv, 69978, Israel}
\newcommand{\instToho}{Toho University, Funabashi 274-8510, Japan}
\newcommand{\instTohoku}{Department of Physics, Tohoku University, Sendai 980-8578, Japan}
\newcommand{\instTitech}{Tokyo Institute of Technology, Tokyo 152-8550, Japan}
\newcommand{\instTokyoMetropolitan}{Tokyo Metropolitan University, Tokyo 192-0397, Japan}
\newcommand{\instUAS}{Universidad Autonoma de Sinaloa, Sinaloa 80000, Mexico}
\newcommand{\instNapoliUNIV}{Dipartimento di Scienze Fisiche, Universit\`{a} di Napoli Federico II, I-80126 Napoli, Italy}
\newcommand{\instPadovaUNIV}{Dipartimento di Fisica e Astronomia, Universit\`{a} di Padova, I-35131 Padova, Italy}
\newcommand{\instPerugiaUNIV}{Dipartimento di Fisica, Universit\`{a} di Perugia, I-06123 Perugia, Italy}
\newcommand{\instPisaUNIV}{Dipartimento di Fisica, Universit\`{a} di Pisa, I-56127 Pisa, Italy}
\newcommand{\instRomaTreUNIV}{Dipartimento di Matematica e Fisica, Universit\`{a} di Roma Tre, I-00146 Roma, Italy}
\newcommand{\instTorinoUNIV}{Dipartimento di Fisica, Universit\`{a} di Torino, I-10125 Torino, Italy}
\newcommand{\instTriesteUNIV}{Dipartimento di Fisica, Universit\`{a} di Trieste, I-34127 Trieste, Italy}
\newcommand{\instMontreal}{Universit\'{e} de Montr\'{e}al, Physique des Particules, Montr\'{e}al, Qu\'{e}bec, H3C 3J7, Canada}
\newcommand{\instIJCLab}{Universit\'{e} Paris-Saclay, CNRS/IN2P3, IJCLab, 91405 Orsay, France}
\newcommand{\instIPHC}{Universit\'{e} de Strasbourg, CNRS, IPHC, UMR 7178, 67037 Strasbourg, France}
\newcommand{\instAdelaide}{Department of Physics, University of Adelaide, Adelaide, South Australia 5005, Australia}
\newcommand{\instBonn}{University of Bonn, 53115 Bonn, Germany}
\newcommand{\instUBC}{University of British Columbia, Vancouver, British Columbia, V6T 1Z1, Canada}
\newcommand{\instCincinnati}{University of Cincinnati, Cincinnati, Ohio 45221, U.S.A.}
\newcommand{\instFlorida}{University of Florida, Gainesville, Florida 32611, U.S.A.}
\newcommand{\instHawaii}{University of Hawaii, Honolulu, Hawaii 96822, U.S.A.}
\newcommand{\instHeidelberg}{University of Heidelberg, 68131 Mannheim, Germany}
\newcommand{\instLjubljanaUniLJ}{Faculty of Mathematics and Physics, University of Ljubljana, 1000 Ljubljana, Slovenia}
\newcommand{\instLouisville}{University of Louisville, Louisville, Kentucky 40292, U.S.A.}
\newcommand{\instMalaya}{National Centre for Particle Physics, University Malaya, 50603 Kuala Lumpur, Malaysia}
\newcommand{\instLjubljanaUM}{University of Maribor, 2000 Maribor, Slovenia}
\newcommand{\instMelbourne}{School of Physics, University of Melbourne, Victoria 3010, Australia}
\newcommand{\instMississippi}{University of Mississippi, University, Mississippi 38677, U.S.A.}
\newcommand{\instUOM}{University of Miyazaki, Miyazaki 889-2192, Japan}
\newcommand{\instPittsburgh}{University of Pittsburgh, Pittsburgh, Pennsylvania 15260, U.S.A.}
\newcommand{\instUSTC}{University of Science and Technology of China, Hefei 230026, China}
\newcommand{\instSAlabama}{University of South Alabama, Mobile, Alabama 36688, U.S.A.}
\newcommand{\instSCarolina}{University of South Carolina, Columbia, South Carolina 29208, U.S.A.}
\newcommand{\instSydney}{School of Physics, University of Sydney, New South Wales 2006, Australia}
\newcommand{\instUTokyo}{Department of Physics, University of Tokyo, Tokyo 113-0033, Japan}
\newcommand{\instIPMU}{Kavli Institute for the Physics and Mathematics of the Universe (WPI), University of Tokyo, Kashiwa 277-8583, Japan}
\newcommand{\instVictoria}{University of Victoria, Victoria, British Columbia, V8W 3P6, Canada}
\newcommand{\instVPI}{Virginia Polytechnic Institute and State University, Blacksburg, Virginia 24061, U.S.A.}
\newcommand{\instWayneState}{Wayne State University, Detroit, Michigan 48202, U.S.A.}
\newcommand{\instYamagata}{Yamagata University, Yamagata 990-8560, Japan}
\newcommand{\instYerevan}{Alikhanyan National Science Laboratory, Yerevan 0036, Armenia}
\newcommand{\instYonsei}{Yonsei University, Seoul 03722, South Korea}
\newcommand{\instZZU}{Zhengzhou University, Zhengzhou 450001, China}
\affiliation{\instCPPM}
\affiliation{\instBeihang}
\affiliation{\instBNL}
\affiliation{\instBINP}
\affiliation{\instCMU}
\affiliation{\instCinvestavIPN}
\affiliation{\instPrague}
\affiliation{\instChiangMai}
\affiliation{\instChiba}
\affiliation{\instChonnam}
\affiliation{\instConacyt}
\affiliation{\instDESY}
\affiliation{\instDuke}
\affiliation{\instITAR}
\affiliation{\instRomaENEA}
\affiliation{\instEri}
\affiliation{\instJuelich}
\affiliation{\instFuJen}
\affiliation{\instFudan}
\affiliation{\instGoettingen}
\affiliation{\instGifu}
\affiliation{\instSOKENDAI}
\affiliation{\instGyeongsang}
\affiliation{\instHanyang}
\affiliation{\instKEK}
\affiliation{\instJPARC}
\affiliation{\instHSE}
\affiliation{\instIISER}
\affiliation{\instIITBhubaneswar}
\affiliation{\instIITGuwahati}
\affiliation{\instIITHyderabad}
\affiliation{\instIITMadras}
\affiliation{\instIndiana}
\affiliation{\instIHEPRussia}
\affiliation{\instHEPHYVienna}
\affiliation{\instHiroshima}
\affiliation{\instIHEPChina}
\affiliation{\instIPP}
\affiliation{\instIOP}
\affiliation{\instIFIC}
\affiliation{\instFrascati}
\affiliation{\instNapoliINFN}
\affiliation{\instPadovaINFN}
\affiliation{\instPerugiaINFN}
\affiliation{\instPisaINFN}
\affiliation{\instRomaINFN}
\affiliation{\instRomaTreINFN}
\affiliation{\instTorinoINFN}
\affiliation{\instTriesteINFN}
\affiliation{\instJAEA}
\affiliation{\instMainz}
\affiliation{\instGiessen}
\affiliation{\instKarlsruhe}
\affiliation{\instISU}
\affiliation{\instKitasato}
\affiliation{\instKISTI}
\affiliation{\instKoreaUnivKU}
\affiliation{\instKSU}
\affiliation{\instKyungpook}
\affiliation{\instLPI}
\affiliation{\instLNNU}
\affiliation{\instLMU}
\affiliation{\instLuther}
\affiliation{\instMNITJaipur}
\affiliation{\instMPP}
\affiliation{\instMPGHLL}
\affiliation{\instMcGill}
\affiliation{\instMEPhI}
\affiliation{\instNagoya}
\affiliation{\instNagoyaIAR}
\affiliation{\instNagoyaKMI}
\affiliation{\instNaraWu}
\affiliation{\instNTUTaiwan}
\affiliation{\instNUUTaiwan}
\affiliation{\instKrakow}
\affiliation{\instNiigata}
\affiliation{\instNSU}
\affiliation{\instOkinawa}
\affiliation{\instOsakaCity}
\affiliation{\instRCNP}
\affiliation{\instPNNL}
\affiliation{\instPanjab}
\affiliation{\instPanjabPAU}
\affiliation{\instRIKENMSL}
\affiliation{\instSeoul}
\affiliation{\instSPU}
\affiliation{\instSoochow}
\affiliation{\instSoongsil}
\affiliation{\instLjubljanaJSI}
\affiliation{\instKyiv}
\affiliation{\instTata}
\affiliation{\instTUM}
\affiliation{\instTelAviv}
\affiliation{\instToho}
\affiliation{\instTohoku}
\affiliation{\instTitech}
\affiliation{\instTokyoMetropolitan}
\affiliation{\instUAS}
\affiliation{\instNapoliUNIV}
\affiliation{\instPadovaUNIV}
\affiliation{\instPerugiaUNIV}
\affiliation{\instPisaUNIV}
\affiliation{\instRomaTreUNIV}
\affiliation{\instTorinoUNIV}
\affiliation{\instTriesteUNIV}
\affiliation{\instMontreal}
\affiliation{\instIJCLab}
\affiliation{\instIPHC}
\affiliation{\instAdelaide}
\affiliation{\instBonn}
\affiliation{\instUBC}
\affiliation{\instCincinnati}
\affiliation{\instFlorida}
\affiliation{\instHawaii}
\affiliation{\instHeidelberg}
\affiliation{\instLjubljanaUniLJ}
\affiliation{\instLouisville}
\affiliation{\instMalaya}
\affiliation{\instLjubljanaUM}
\affiliation{\instMelbourne}
\affiliation{\instMississippi}
\affiliation{\instUOM}
\affiliation{\instPittsburgh}
\affiliation{\instUSTC}
\affiliation{\instSAlabama}
\affiliation{\instSCarolina}
\affiliation{\instSydney}
\affiliation{\instUTokyo}
\affiliation{\instIPMU}
\affiliation{\instVictoria}
\affiliation{\instVPI}
\affiliation{\instWayneState}
\affiliation{\instYamagata}
\affiliation{\instYerevan}
\affiliation{\instYonsei}
\affiliation{\instZZU}
  \author{F.~Abudin{\'e}n}\affiliation{\instTriesteINFN} 
  \author{I.~Adachi}\affiliation{\instKEK}\affiliation{\instSOKENDAI} 
  \author{R.~Adak}\affiliation{\instFudan} 
  \author{K.~Adamczyk}\affiliation{\instKrakow} 
  \author{P.~Ahlburg}\affiliation{\instBonn} 
  \author{J.~K.~Ahn}\affiliation{\instKoreaUnivKU} 
  \author{H.~Aihara}\affiliation{\instUTokyo} 
  \author{N.~Akopov}\affiliation{\instYerevan} 
  \author{A.~Aloisio}\affiliation{\instNapoliUNIV}\affiliation{\instNapoliINFN} 
  \author{F.~Ameli}\affiliation{\instRomaINFN} 
% ...
\collaboration{Belle II Collaboration}

\begin{abstract}
We report updated measurements of branching fractions~($\mathcal{B}$) and CP-violating charge asymmetries~($\mathcal{A_{\rm CP}}$) for charmless $B$ decays at Belle II, which operates on or near the $\Upsilon(4S)$ resonance at the SuperKEKB asymmetric energy $e^{+}e^{-}$ collider. We use samples of 2019 and 2020 data corresponding to $62.8\,\si{fb^{-1}}$ of integrated luminosity. The samples are analysed using two-dimensional fits in $\Delta E$ and $M_{\it bc}$ to determine signal yields of approximately 568, 103, and 115~decays for the channels \mbox{$B^0 \to K^+\pi^-$}, \mbox{$B^+ \to \PKzS\pi^+$}, and \mbox{$B^0 \to \pi^+\pi^-$}, respectively. We obtain the following determinations of branching fractions,
\begin{center}
$\mathcal{B}(B^0 \to K^+\pi^-) = [18.0 \pm 0.9(\rm stat) \pm 0.9(\rm syst)]\times 10^{-6}$,
\end{center}

\begin{center}
$\mathcal{B}(B^+ \to K^0\pi^+) = [21.4 ^{+2.3}_{-2.2}(\rm stat) \pm 1.6(\rm syst)]\times 10^{-6}$,
\end{center}

\begin{center}
$\mathcal{B}(B^0 \to \pi^+\pi^-) = [5.8 \pm 0.7(\rm stat) \pm 0.3(\rm syst)]\times 10^{-6}$,
\end{center}

and CP-violating rate asymmetries,

\begin{center}
$\mathcal{A_{\rm CP}}(B^0 \to K^+\pi^-) = -0.16 \pm 0.05(\rm stat) \pm 0.01(\rm syst)$, \\
\end{center}
\begin{center}
$\mathcal{A_{\rm CP}}(B^+ \to \PKzS\pi^+) = -0.01 \pm 0.08(\rm stat) \pm 0.05(\rm syst)$.
\end{center}
The results are compatible with known determinations and contribute important information to an early assessment of Belle II detector performance.

\end{abstract}

\maketitle
{\renewcommand{\thefootnote}{\fnsymbol{footnote}}}
\setcounter{footnote}{0}

\clearpage


\section{Introduction and Motivation}

\begin{figure}[htb]
 \centering
 \includegraphics[width=0.5\textwidth]{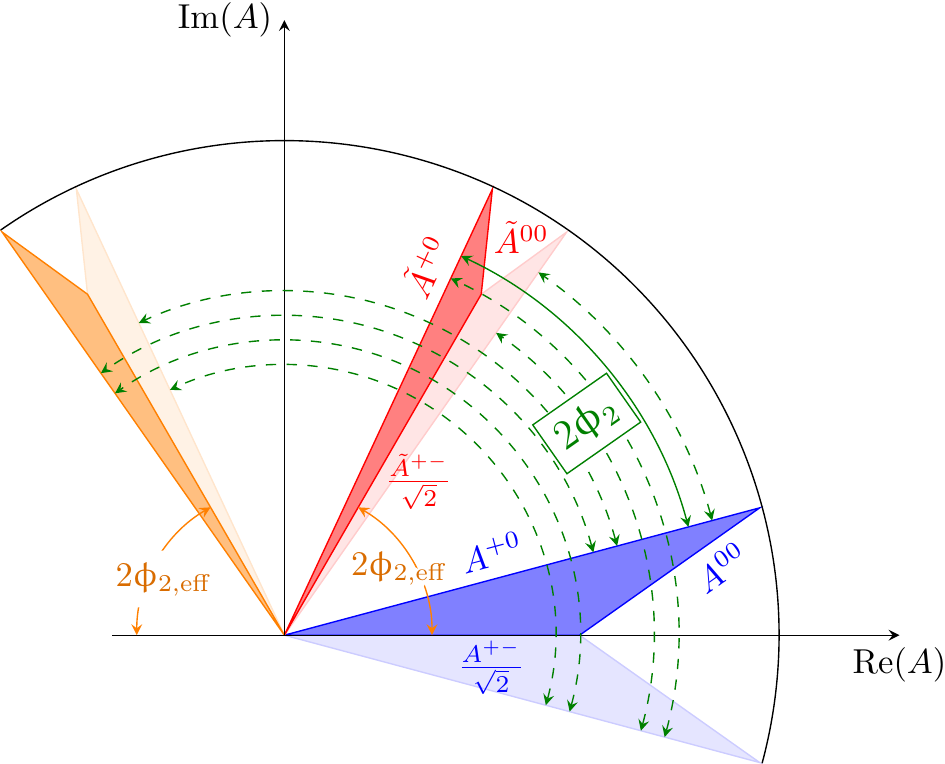}
 \caption{Geometrical representation of the isospin triangular relations in the complex plane of $\PB^{i+j}\to \Ph^i\Ph^j$ amplitudes. The blue and the red shaded areas correspond to the isospin triangles. The angle between the CP-conjugate amplitudes $A^{+-}$ and $\tilde{A}^{+-}$ corresponds to twice the weak phase $\phi_{2, {\rm eff}}/\alpha_{\rm eff}$ (orange arrows). The angle between the CP-conjugate amplitudes $A^{+0}$ and $\tilde{A}^{+0}$ corresponds to twice the CKM angle $\phi_2/\alpha$ (green solid arrow). The triangles with lighter shade represent the mirror solutions allowed by discrete ambiguities, with the corresponding values for $\alpha$ represented by the green dashed lines.}
 \label{fig:isospin}
\end{figure}

The study of charmless $B$ decays is a keystone of the worldwide flavor program. Processes mediated by $b\to u\bar{u}d$ transitions offer direct access to the unitarity angle $\phi_2/\alpha$ and probe contributions of non-standard-model dynamics in loops.  However, reliable extraction of weak phases and unambiguous interpretation of measurements involving loop amplitudes are spoiled by large hadronic uncertainties,  which are rarely tractable in perturbative calculations. Appropriately chosen combinations of measurements from decay modes related by flavor symmetries are used to reduce the impact of such unknowns. An especially  fruitful approach consists in combining measurements of decays related by isospin symmetries. For instance, a combined analysis of all isospin partners of $\B \to \pi\pi$ allows for a direct measurement of $\phi_2/\alpha$ (compare Fig.~\ref{fig:isospin}), rendering it an effective complementary method to the indirect determination~\cite{Charles:2017}. A similar approach has been proposed to address the so-called $K^+\pi^-$ puzzle, a long standing anomaly associated with the significant difference between direct CP-violating asymmetries observed in $B^0 \to K^+\pi^-$ and $B^+ \to K^+\pi^0$ decays ~\cite{Gronau:1999}. The asymmetries are expected to be equal at the leading order, as the two decays differ only in the spectator quark.
The isospin-sum rule
\begin{equation}
I_{K\pi} = \mathcal{A}_{K^+\pi^-} + \mathcal{A}_{K^0\pi^+} \cdot \frac{\mathcal{B}(K^0\pi^+)}{\mathcal{B}(K^+\pi^-)}\frac{\tau_{B^0}}{\tau_{B^+}} - 2\mathcal{A}_{K^+\pi^0} \cdot \frac{\mathcal{B}(K^+\pi^0)}{\mathcal{B}(K^+\pi^-)}\frac{\tau_{B^0}}{\tau_{B^+}} -2\mathcal{A}_{K^0\pi^0}\cdot\frac{\mathcal{B}(K^0\pi^0)}{\mathcal{B}(K^+\pi^-)}
\end{equation}
properly accounts for sub-leading amplitudes by combining the branching fractions ($\mathcal{B}$) and direct CP-asymmetries ($\mathcal{A_{\rm CP}}$) of $B$ decays to all four final states $K^+\pi^-$, $K^0\pi^+$, $K^+\pi^0$ and $K^0\pi^0$. This relationship offers a good test of the standard model (SM), which predicts $I_{K\pi}\approx0$ in the limit of isospin symmetry and assuming no electroweak penguin contributions, with an uncertainty of $\mathcal{O}$(1$\%$)~\cite{Gronau:2005, Browder:2009, Gershon:2007, Bell:2015}. Belle II has the unique capability of studying jointly, and within a consistent experimental environment, all relevant final states.

The Belle II experiment, complete with its vertex detector, started its collision operations on March 2019 and is currently taking data. The sample of electron-positron collisions used in this work corresponds to an integrated luminosity of
62.8 fb$^{-1}$ at the $\Upsilon(4{\rm S})$ resonance. This document presents improved measurements of branching fraction and charge-parity violating charge asymmetries in \mbox{$B^0 \to K^+\pi^-$}, \mbox{$B^+ \to \PKzS\pi^+$}, and \mbox{$B^0 \to \pi^+\pi^-$} decays that improve and supersede previous Belle II results~\cite{Abudinen:2020ICHEP}.
Besides roughly doubling the size of the data set, the sample-composition determinations are now based on simultaneous fits to the energy-difference and beam-constrained-mass distributions. In addition, we adopt a refined treatment of peaking-background contributions and systematic uncertainties.

All analysis procedures are first developed and finalized in simulated data. We test the analysis on the subset of data used in our previous report~\cite{Abudinen:2020ICHEP}, corresponding to 55$\%$ of the total sample, prior to the application on the full data set. A signal selection is applied to suppress the major sources of backgrounds, building on previous work~\cite{Abudinen:2020ICHEP}. Fits then determine the sample composition in terms of signal, background from $e^{+}e^{-} \to q\bar{q} $ continuum events, where $q$ indicates any quark of the first or second family ($u$, $d$, $s$, and $c$), and background from non-signal $B$-decays, using the following observables:
\begin{itemize}
    \item the energy difference $\Delta E \equiv E^{*}_{B} - \sqrt{s}/2$ between the total energy of the reconstructed $B$ candidate and half of the collision energy, both in the $\Upsilon(4{\rm S})$ frame;
    \item the beam-energy-constrained mass $M_{\rm bc} \equiv \sqrt{s/(4c^4) - (p^{*}_B/c)^2}$, which is the invariant mass of the $B$ candidate where the $B$ energy is replaced by the (more precisely known) half of the center-of-mass collision energy. 
\end{itemize}

Charge-conjugate processes are implied in what follows except when otherwise stated.

\section{The Belle~II detector}
Belle~II is a $4\pi$ particle-physics spectrometer~\cite{Kou:2018nap, Abe:2010sj}, designed to reconstruct the products of electron-positron collisions produced by the SuperKEKB asymmetric-energy collider~\cite{Akai:2018mbz}, located at the KEK laboratory in Tsukuba, Japan. Belle~II comprises several subdetectors arranged around the interaction space-point in a cylindrical geometry. The innermost subdetector is the vertex detector, which uses position-sensitive silicon layers to determine the trajectories of charged particles (tracks) in the vicinity of the interaction region in order to extrapolate the decay positions of their long-lived parent particles. The vertex detector includes two inner layers of silicon pixel sensors and four outer layers of silicon microstrip sensors. The second pixel layer is currently incomplete and covers only a small portion of azimuthal angle. Charged-particle momenta and charges are measured by a large-radius, helium-ethane, small-cell central drift chamber, which also offers charged-particle-identification information through a measurement of particles' energy-loss by specific ionization. A Cherenkov time-of-propagation detector surrounding the chamber provides charged-particle identification in the central detector volume, supplemented by proximity-focusing, aerogel, ring-imaging Cherenkov detectors in the forward regions. A CsI(Tl)-crystal electromagnetic calorimeter allows for energy measurements of electrons and photons.  A solenoid surrounding the calorimeter generates a uniform axial 1.5\,T magnetic field filling its inner volume. In the barrel region layers of plastic scintillators and resistive-plate chambers, interspersed between the
magnetic flux-return iron plates, allow for identification of $K^0_{\rm L}$ and muons.
The subdetectors most relevant for this work are the silicon vertex detector, the tracking drift chamber, the particle-identification detectors, and the electromagnetic calorimeter.

\section{Data and simulation}
We use all 2019--2020 $\Upsilon(4{\rm S})$ good-quality data collected up to July 1, 2020 and corresponding to an integrated luminosity of $62.8\,\si{fb^{-1}}$. All events are required to satisfy loose data-skim selection criteria, based on total energy and charged-particle multiplicity in the event, targeted at reducing sample sizes to a manageable level with negligible impact on signal efficiency.
We use generic simulated data to optimize the event selection and compare the distributions observed in experimental data with expectations. We use signal-only simulated data to model relevant signal features for fits and determine selection efficiencies. 
Generic simulation consists of Monte Carlo samples that include $\en \ep \to  B^0\overline{B}^0$, $B^+B^-$, $u\bar{u}$, $d\bar{d}$, $c\bar{c}$, and $s\bar{s}$ processes in realistic proportions and corresponding in size to approximately ten times the $\Upsilon(4{\rm S})$ data. In addition, $2\times 10^6$ signal-only events are generated for each signal mode~\cite{Ryd:2005zz}.
All data are processed using the Belle~II analysis software~\cite{Kuhr:2018lps}.

\section{Reconstruction and baseline selection}
We form final-state particle candidates by applying loose baseline selection criteria and then combine candidates in kinematic fits consistent with the topologies of the desired decays to reconstruct intermediate states and $B$ candidates.

We reconstruct charged pion and kaon candidates by starting from the most inclusive charged-particle classes and by requiring fiducial criteria that restrict them to the full polar-angle acceptance in the central drift chamber ($\SI{17}{\degree}<\theta<\SI{150}{\degree}$) and to loose ranges of displacement from the nominal interaction space-point (radial displacement $|dr|<\SI{0.5}{cm}$ and longitudinal displacement $|dz|<\SI{3}{cm}$) to reduce beam-background-induced tracks, which preferentially do not originate from the interaction region. For $K_{\rm S}^0$ reconstruction, we use pairs of oppositely charged particles with a dipion-mass consistent with a $K_{\rm S}^0$ and a successful vertex fit. To reduce combinatorial background, we apply additional requirements, dependent on $K_{\rm S}^0$~momentum, on the distance between trajectories of the two charged-pion candidates, the $K^0_{\rm S}$~flight distance, and the angle between the pion-pair momentum and the direction of the $K^0_{\rm S}$ candidate.
The resulting $K^+$, $\pi^+$, and $\PKzS$ candidates are combined through simultaneous kinematic fits of the entire decay chain into each of our target signal channels, consistent with the desired topology.
In addition, we reconstruct the vertex of the accompanying tag-side $B$ mesons using  all tracks on the tag-side and identify the flavor, which is used as input to the continuum-background discriminator, using a category-based flavor tagger~\cite{Abudinen:2018}.

\section{Continuum suppression}
The main challenge in reconstructing significant charmless signals is the large contamination from continuum background. We use a binary boosted decision-tree classifier that combines non-linearly 39 variables known to provide statistical discrimination between $B$-meson signals and continuum and to be loosely correlated with $\Delta E$ and $M_{\rm bc}$. The variables include quantities associated with event topology, flavor-tagger information, vertex separation and uncertainty information, and kinematic-fit quality information. We train the classifier to identify statistically significant signal and background features using unbiased simulated samples.
We validate the input and output distributions of the classifier by comparing data with simulation using control sample of {$\PBplus\to\APD^{0}(\to \PKp\Pgpm)\,\Pgpp$} decays.

\section{Optimization of the signal selection}
For each channel, we optimize the selection to isolate abundant, low-background signals. We simultaneously vary the selection criteria on continuum-suppression output and charged-particle identification (PID) information to maximize ${\rm S}/\sqrt{{\rm S}+{\rm B}}$, where ${\rm S}$ and ${\rm B}$ are simulated signal and background yields respectively, estimated inside a signal-rich region. The optimal PID requirement corresponds to a kaon (pion) selection efficiency of 74\% (82\%)  with a pion-to-kaon (kaon-to-pion) misidentification rate of 12\% (24\%) in the $\PKp\Pgpm$ ($\Pgpp\Pgpm$) sample.
The resulting signal-candidate multiplicities range from 1.0 to 1.2. In cases with multiple $B$-meson candidates, we choose the candidate randomly to avoid any biases.

\section{Determination of signal yields}
\label{sec:yields}
Signal yields are determined with two-dimensional maximum likelihood fits of the unbinned $\Delta E$ and $M_{\rm bc}$ distributions of candidates restricted to the signal region $M_{\rm bc} > 5.25$\,GeV/$c^2$ and $-0.15 < \Delta E < 0.15$ GeV. Fit models are determined empirically from simulation, with shifts of peak positions and width scale-factors determined in control data. 

In all three channels the signal component is modeled by the sum of a Gaussian and a Crystal Ball function~\cite{Skwarnicki:1986xj} in $\Delta E$ and a single Gaussian in $M_{bc}$.
Peaking backgrounds originating from misidentified final-state particles in the $K^+\pi^-$ and $\pi^+\pi^-$ samples are modeled using the respective signal models, where the relative shifts from the signal peak position in $\Delta E$ are determined by fits to simulated data. In both cases we allow the normalization to float in the fit to data. Non-negligible residual background from $B$ decays other than signal in $\Bz \to K^+\pi^-$ and $B^+ \to \PKzS \pi^+$ is modeled by a one-dimensional kernel density estimator model and a Crystal Ball in $\Delta E$ and $M_{\rm bc}$, respectively. We fix the normalization of these components to values determined from simulation based on known branching fractions. The remaining background, dominated by continuum events, is modeled with a straight line ($\Delta E$) and an ARGUS function ($M_{\rm bc}$), with all other shape parameters and normalizations floating in fits.

\begin{figure}[htb]
 \centering
 \includegraphics[width=0.475\textwidth]{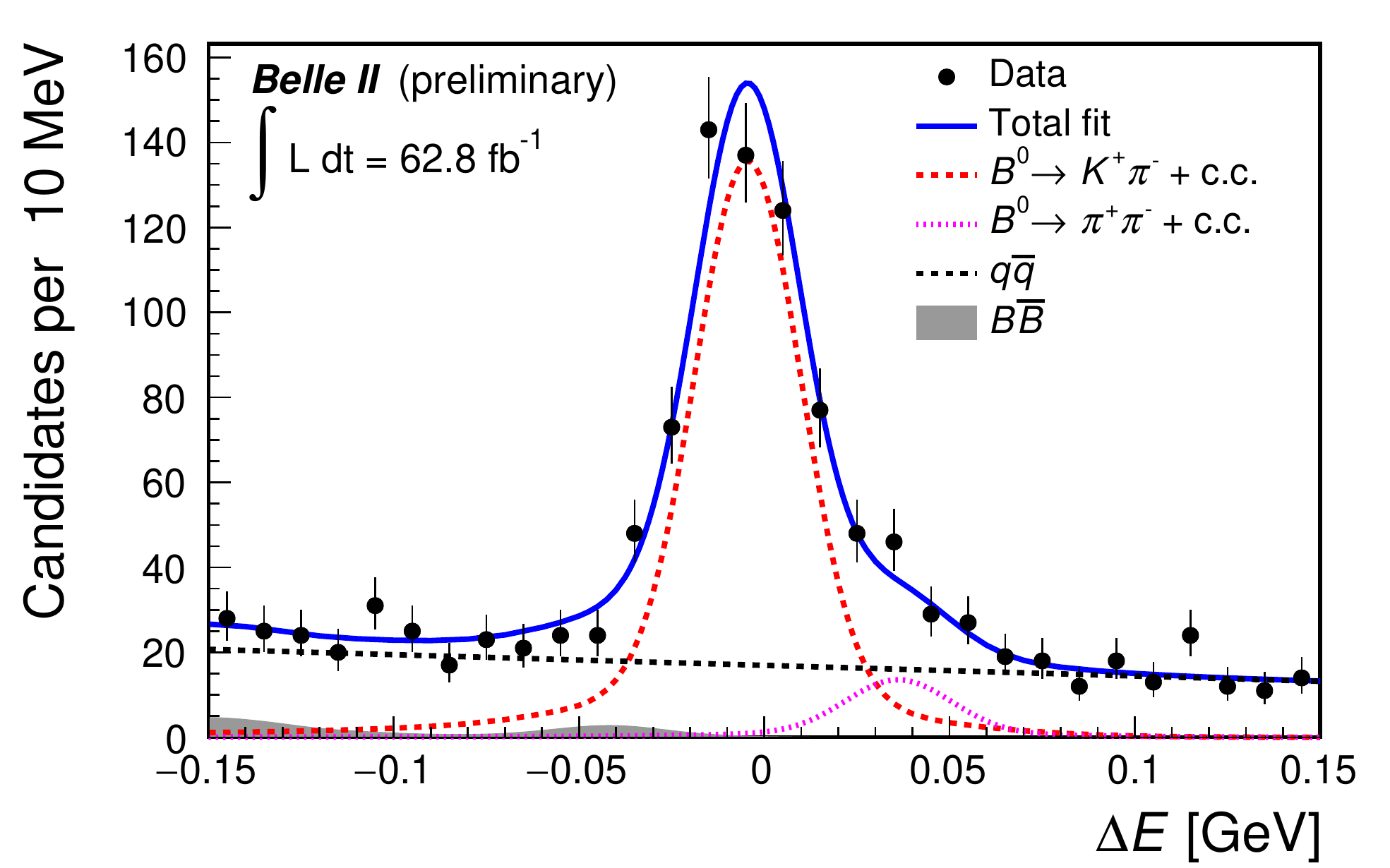}
 \includegraphics[width=0.475\textwidth]{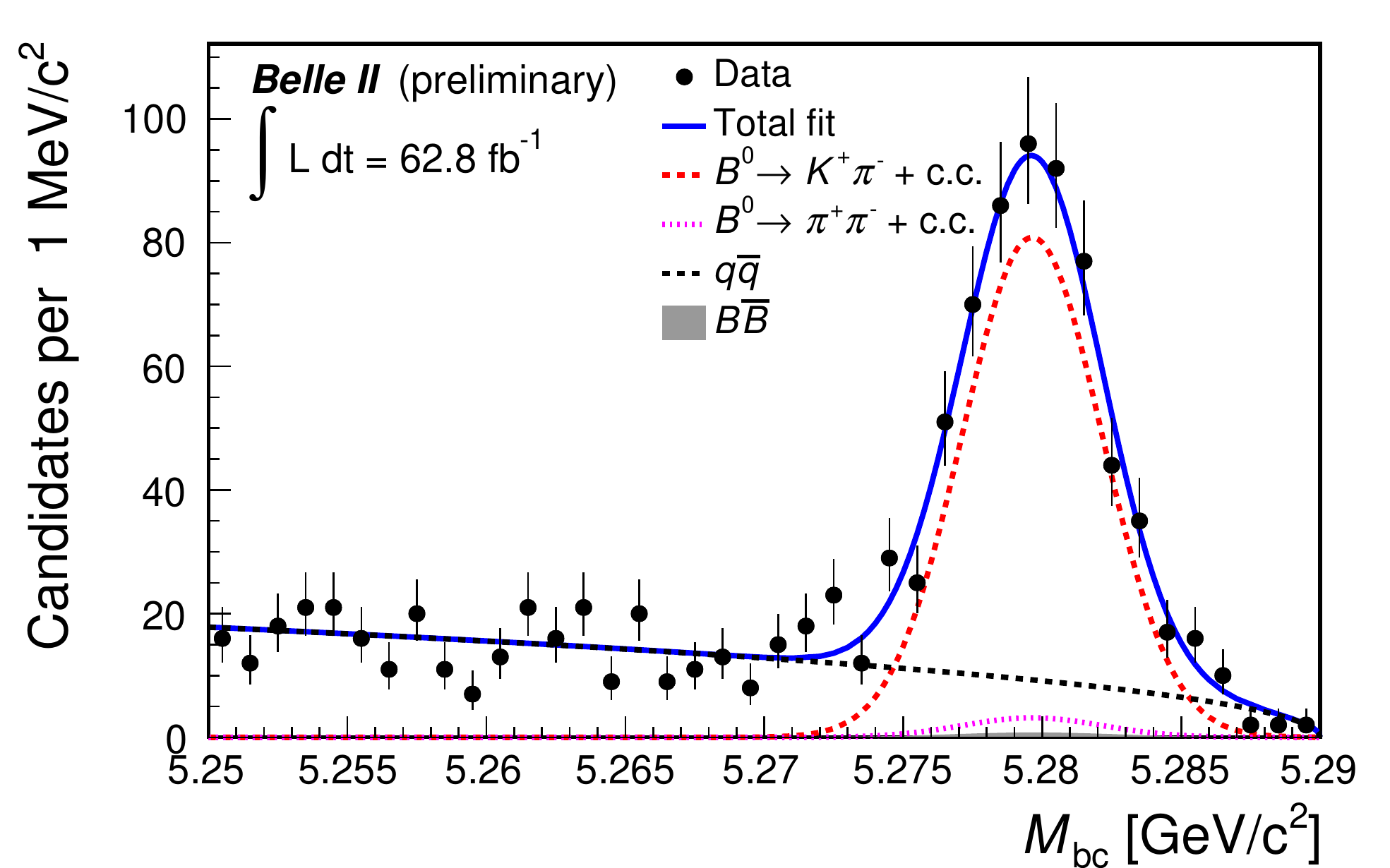}
 \caption{Distributions of $\Delta E$ (left) and $M_{\rm bc}$ (right) for $B^0 \to K^+\pi^-$ candidates reconstructed in 2019--2020 Belle II data, selected with an optimized continuum-suppression and kaon-enriching selection. The distributions are shown in signal-enriched regions of $5.273<M_{\rm bc}<5.286$ GeV/$c^2$ and $-0.04<\Delta E <0.03$ GeV, respectively. Fit projections are overlaid.}
 \label{fig:Kpi_yield}
\end{figure}

\begin{figure}[htb]
 \centering
 \includegraphics[width=0.475\textwidth]{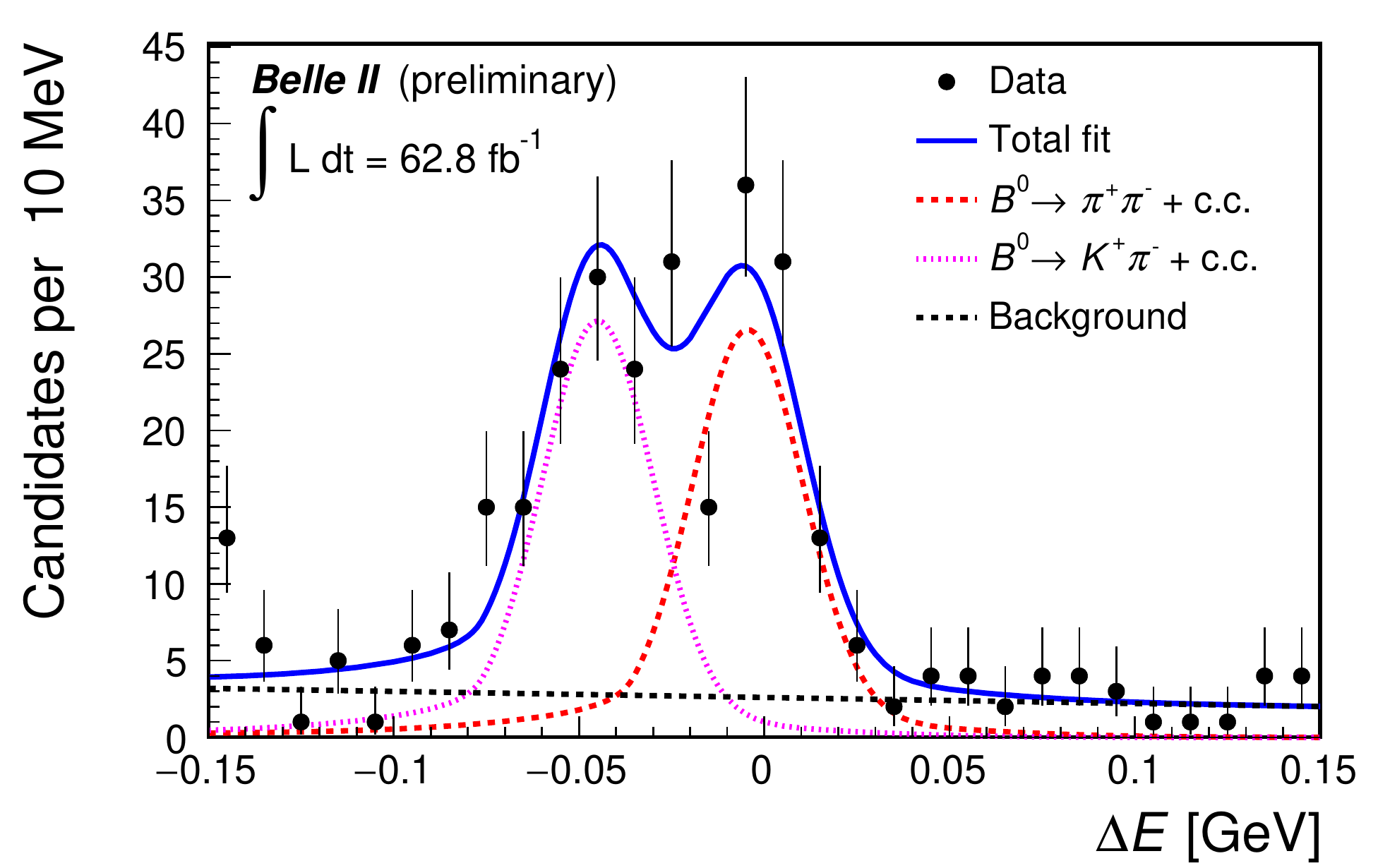}
 \includegraphics[width=0.475\textwidth]{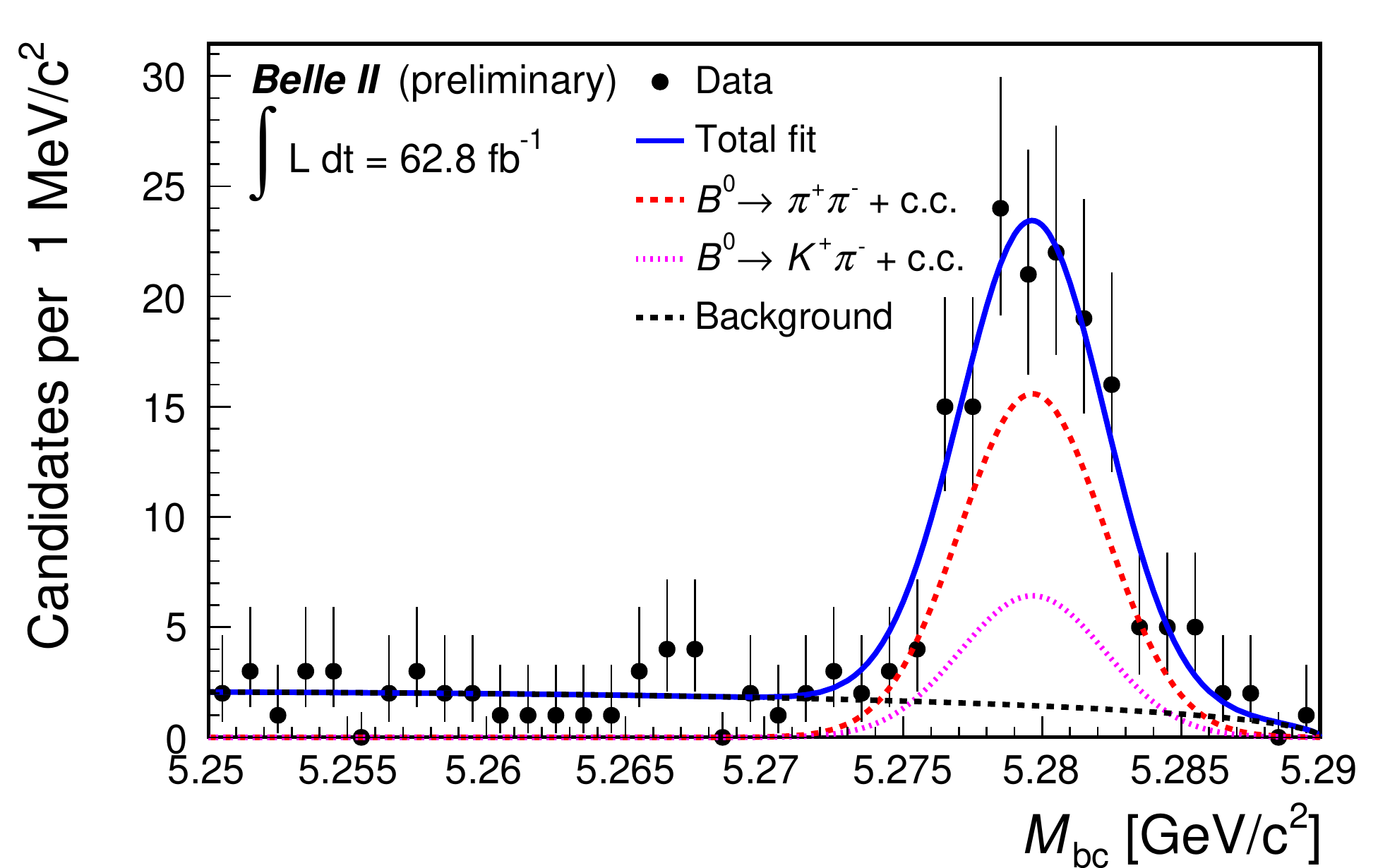}
 \caption{Distributions of $\Delta E$ (left) and $M_{\rm bc}$ (right) for $B^0 \to \pi^+\pi^-$ candidates reconstructed in 2019--2020 Belle II data, selected with an optimized continuum-suppression and pion-enriching selection. The distributions are shown in signal-enriched regions of $5.273<M_{\rm bc}<5.286$ GeV/$c^2$ and $-0.04<\Delta E <0.03$ GeV, respectively. Fit projections are overlaid.}
 \label{fig:pipi_yield}
\end{figure}

\begin{figure}[htb]
 \centering
 \includegraphics[width=0.475\textwidth]{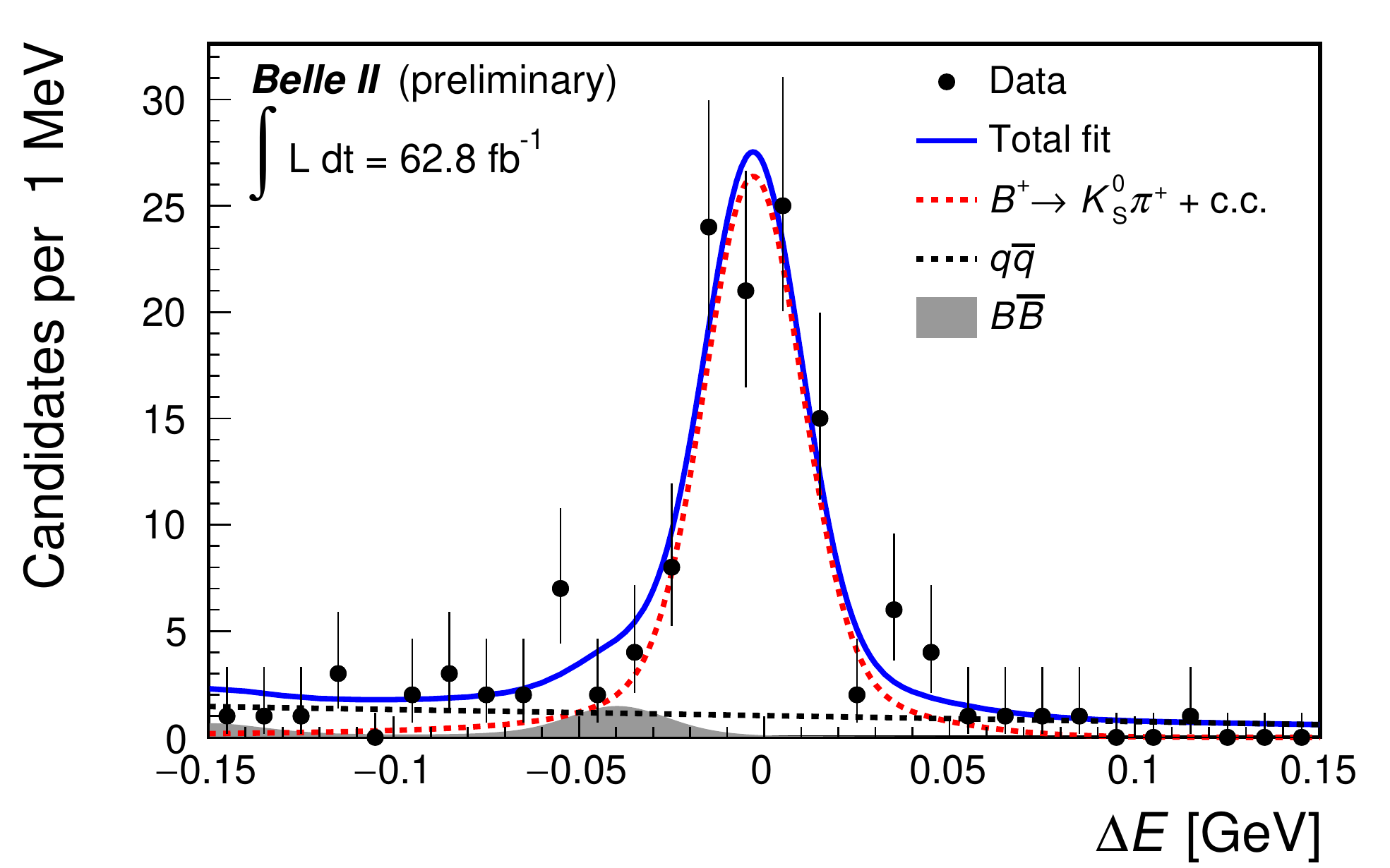}
 \includegraphics[width=0.475\textwidth]{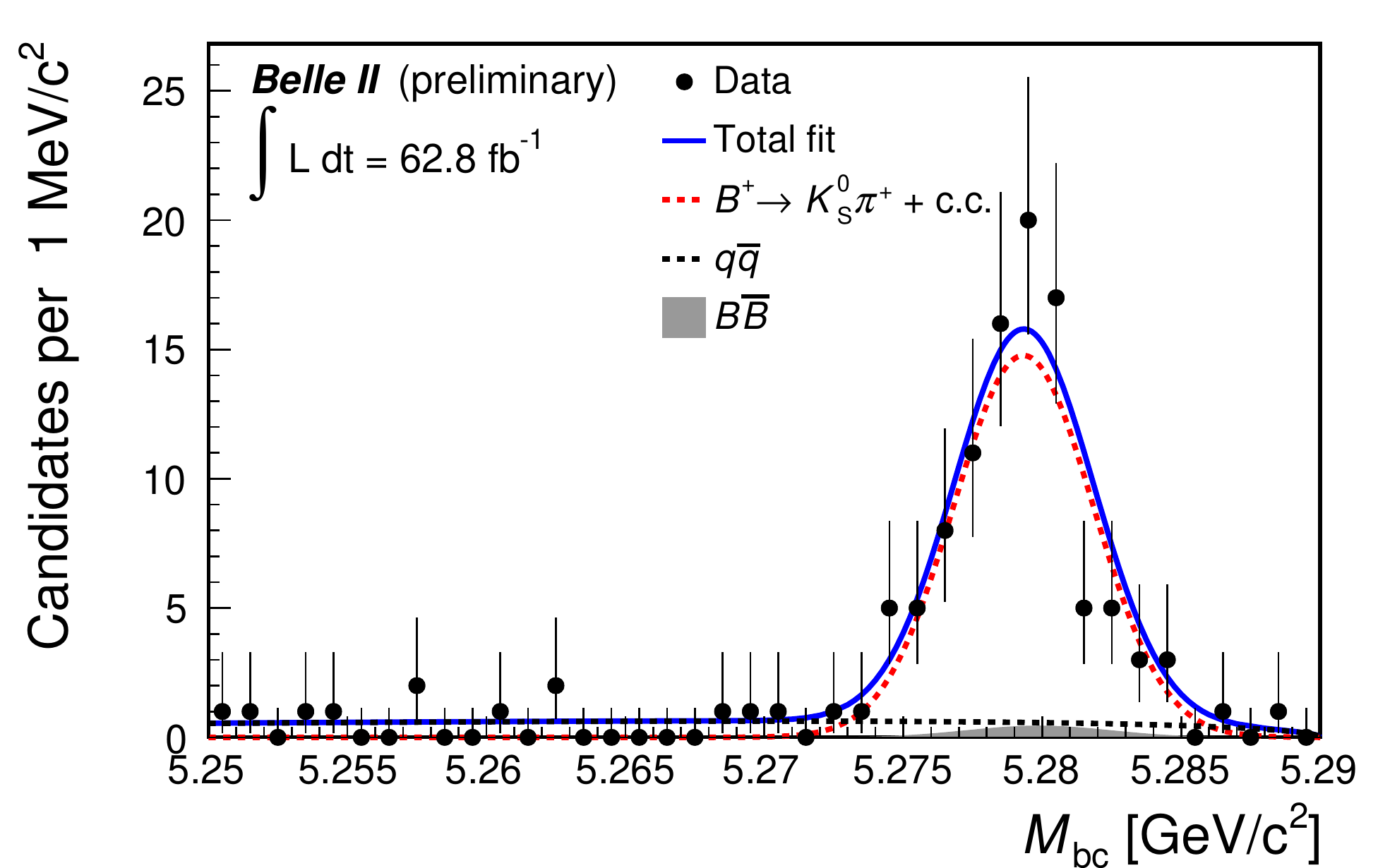}
 \caption{Distributions of $\Delta E$ (left) and $M_{\rm bc}$ (right) for $B^+ \to \PKzS \pi^+$ candidates reconstructed in 2019--2020 Belle II data, selected with an optimized continuum-suppression. The distributions are shown in signal-enriched regions of $5.273<M_{\rm bc}<5.286$ GeV/$c^2$ and $-0.04<\Delta E <0.03$ GeV, respectively. Fit projections are overlaid.}
 \label{fig:Kspi_yield}
\end{figure}

The $\Delta E$ and $M_{\rm bc}$ distributions with fit projections overlaid are shown in Figs.~\ref{fig:Kpi_yield}--\ref{fig:Kspi_yield}. Prominent narrow signals are visible overlapping smooth backgrounds dominated by continuum.

\begin{figure}[htb]
 \centering
 \includegraphics[width=0.475\textwidth]{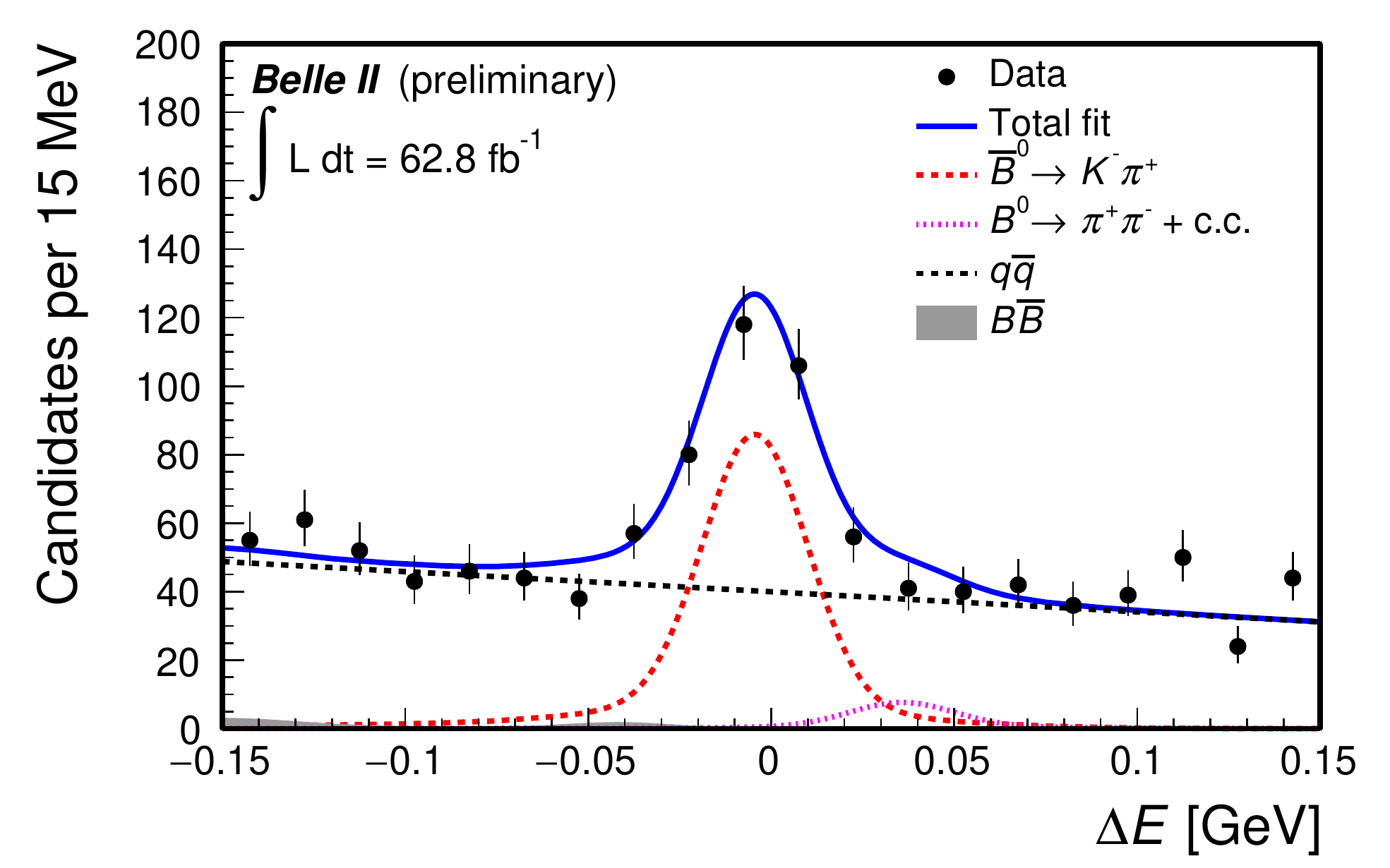}
 \includegraphics[width=0.475\textwidth]{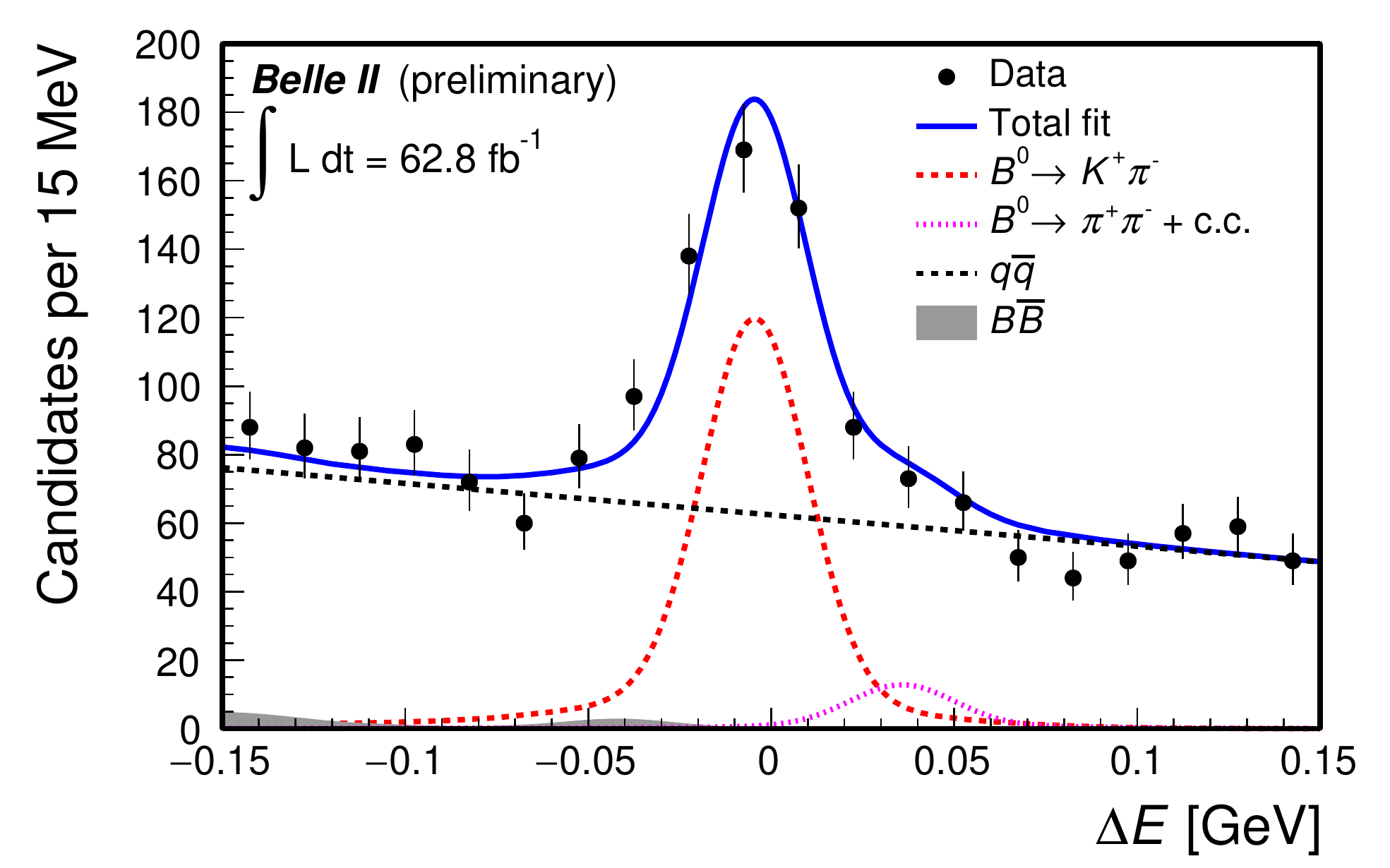}
 \includegraphics[width=0.475\textwidth]{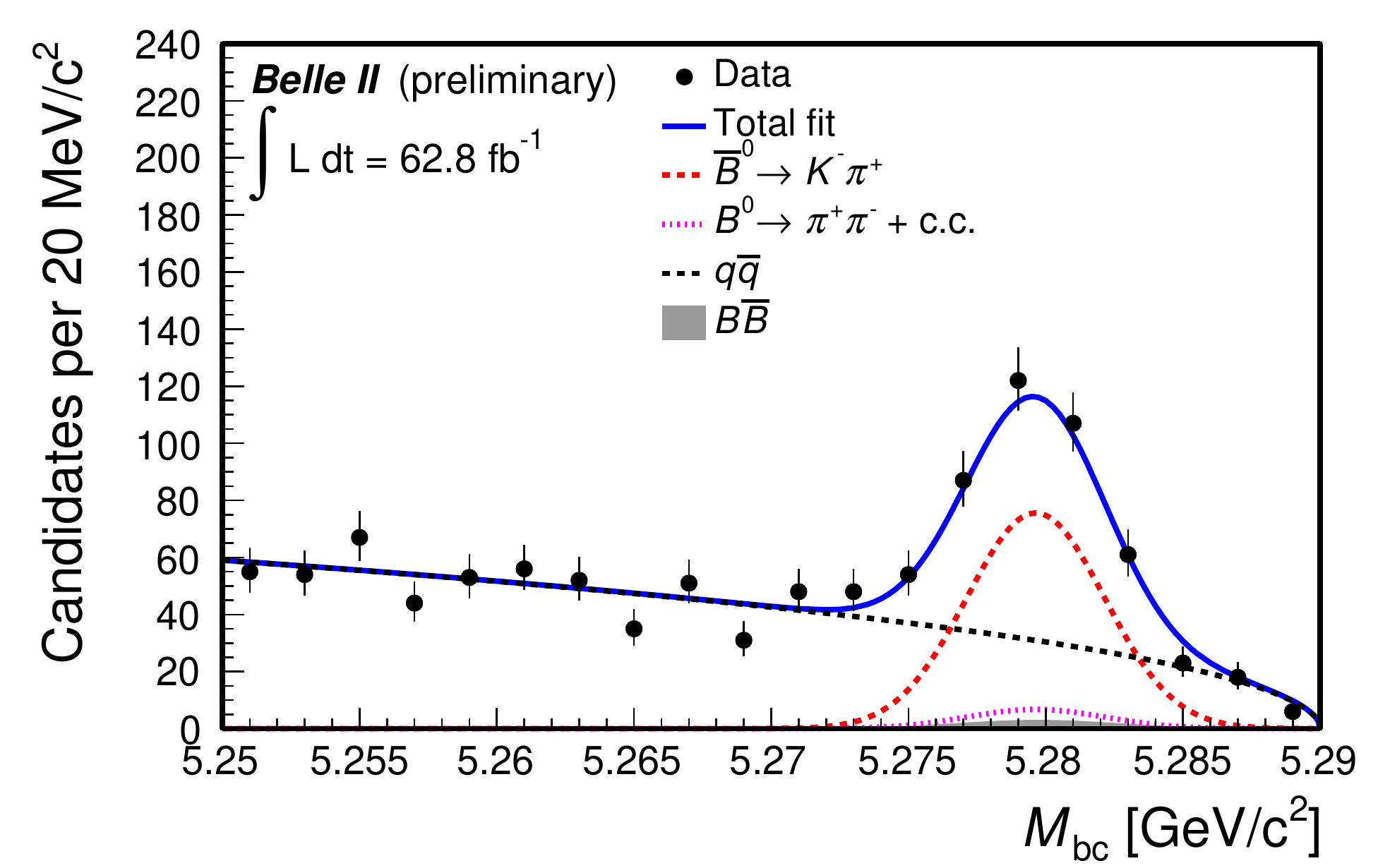}
 \includegraphics[width=0.475\textwidth]{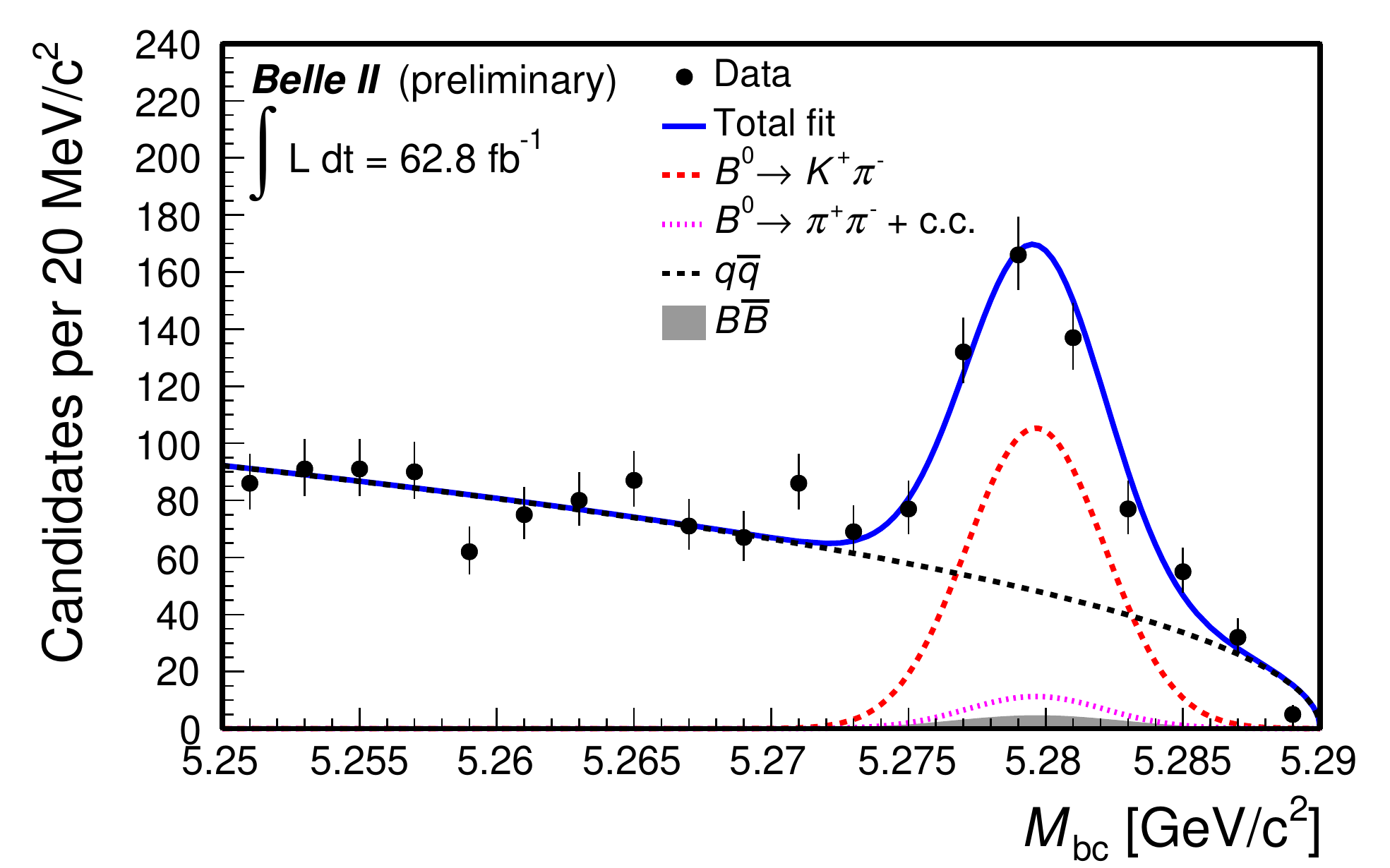} 
 \caption{Distributions of $\Delta E$ (top) and $M_{\rm bc}$ (bottom) for $\Bzb \to K^-\pi^+$ (left) and $\Bz \to K^+\pi^-$ (right) candidates reconstructed 2019--2020 Belle II data, selected with an optimized continuum-suppression. Fit projections are overlaid.}
 \label{fig:Kpi_ACP}
\end{figure}

\begin{figure}[htb]
 \centering
 \includegraphics[width=0.475\textwidth]{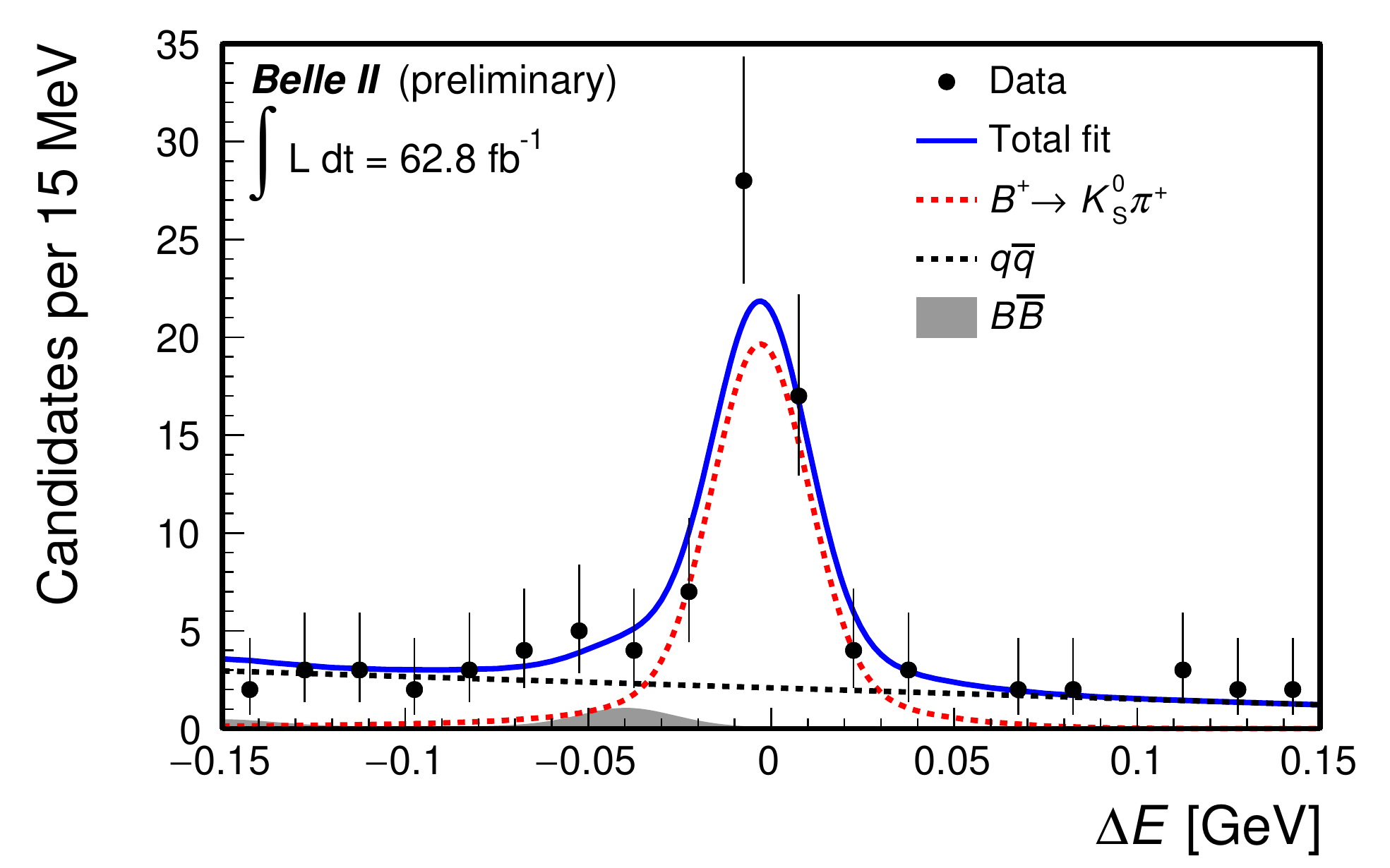}
 \includegraphics[width=0.475\textwidth]{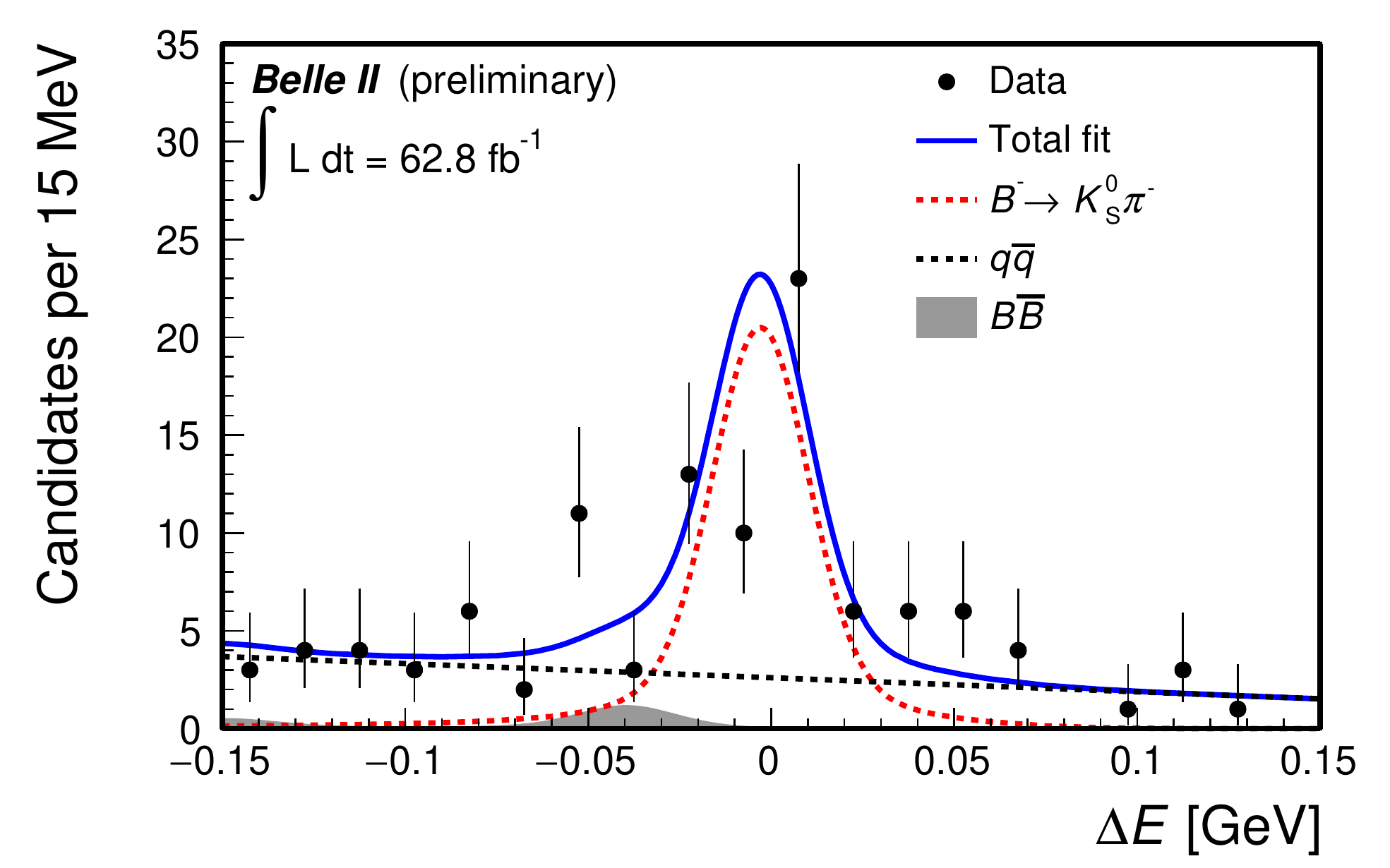}
 \includegraphics[width=0.475\textwidth]{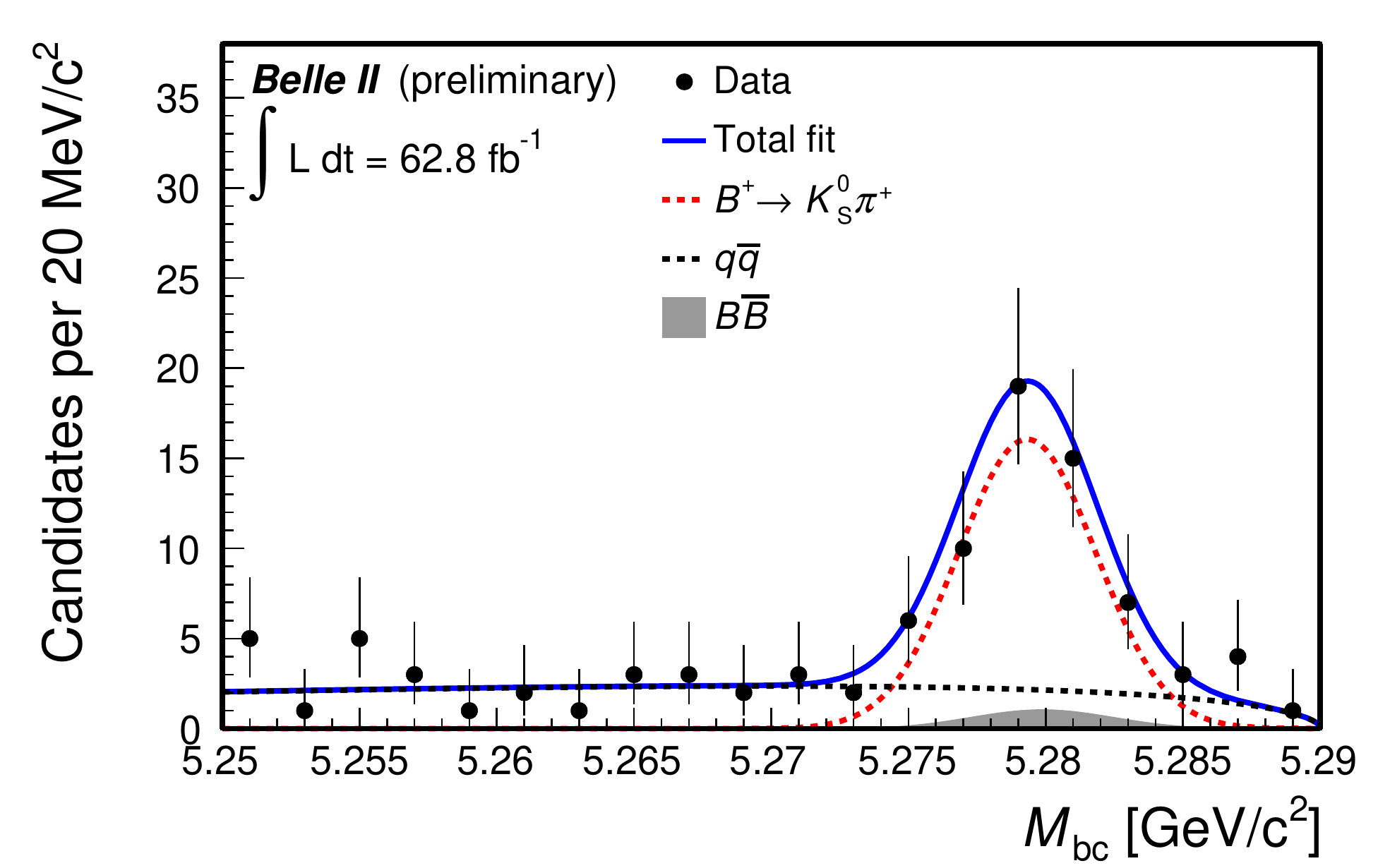}
 \includegraphics[width=0.475\textwidth]{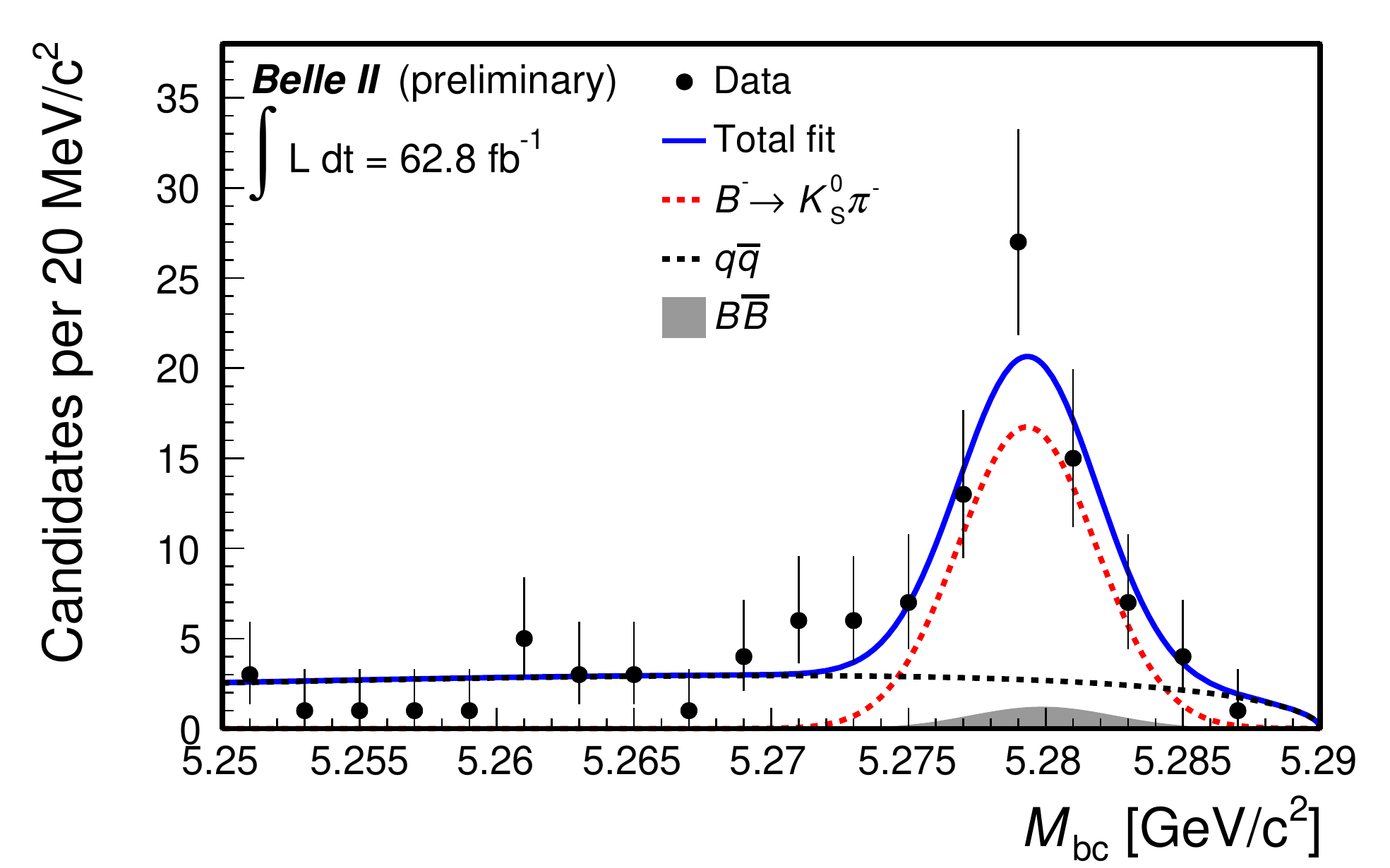} 
 \caption{Distributions of $\Delta E$ (top) and $M_{\rm bc}$ (bottom) for $B^+ \to \PKzS \pi^+$ (left) and $B^- \to \PKzS \pi^-$ (right) candidates reconstructed 2019--2020 Belle II data, selected with an optimized continuum-suppression. Fit projections are overlaid.}
 \label{fig:Kspi_ACP}
\end{figure}

\clearpage
In addition, we use a non-extended likelihood to fit simultaneously the unbinned $\Delta E$ and $M_{\rm bc}$ distributions of bottom and antibottom candidates decaying in flavor-specific final states for measurements of direct CP-violation. We use the same signal and background models as were used for branching-fraction measurements and take the raw partial-decay-rate asymmetry as a fit parameter,
\begin{equation*}
\mathcal{A_{\rm CP}}=\frac{N(b)-N(\bar{b})}{N(b)+N(\bar{b})},
\end{equation*}
where $N$ are signal yields and $b$ ($\bar{b}$) indicates the meson containing a bottom (antibottom) quark. Charge-specific $\Delta E$ and $M_{\rm bc}$ distributions are shown in Figs.~\ref{fig:Kpi_ACP} and \ref{fig:Kspi_ACP} with fit projections overlaid. Table~\ref{tab:ACP} lists the charge-specific yields and resulting raw charge-asymmetries.

 \begin{table}[!ht]
     \centering
 \begin{tabular}{l  r  r  r }
 \hline\hline
 Decay & \multicolumn{1}{c}{~ ~ $B^+$ yield} & \multicolumn{1}{c}{~ $B^-$ yield} & ~ ~ Raw asymmetry  \\\hline
   $B^0 \to K^+\pi^-$    &	$332 \pm 20$  &	$238 \pm 17$ & $-0.16 \pm 0.04$\\ 
   $B^+ \to \PKzS \pi^+$ &	$50 \pm 5 $ &  $53 \pm 6$ & $+0.02\pm 0.08$ \\
   \hline
 \end{tabular}
     \caption{Summary of charge-specific signal yields for the measurement of CP-violating asymmetries in 2019-2020 Belle II data. Only the statistical contributions to the uncertainties are listed.} 
     \label{tab:ACP}
\end{table}{}

\section{Efficiencies and corrections}
\begin{figure}[htb]
 \centering
 \includegraphics[width=0.475\textwidth]{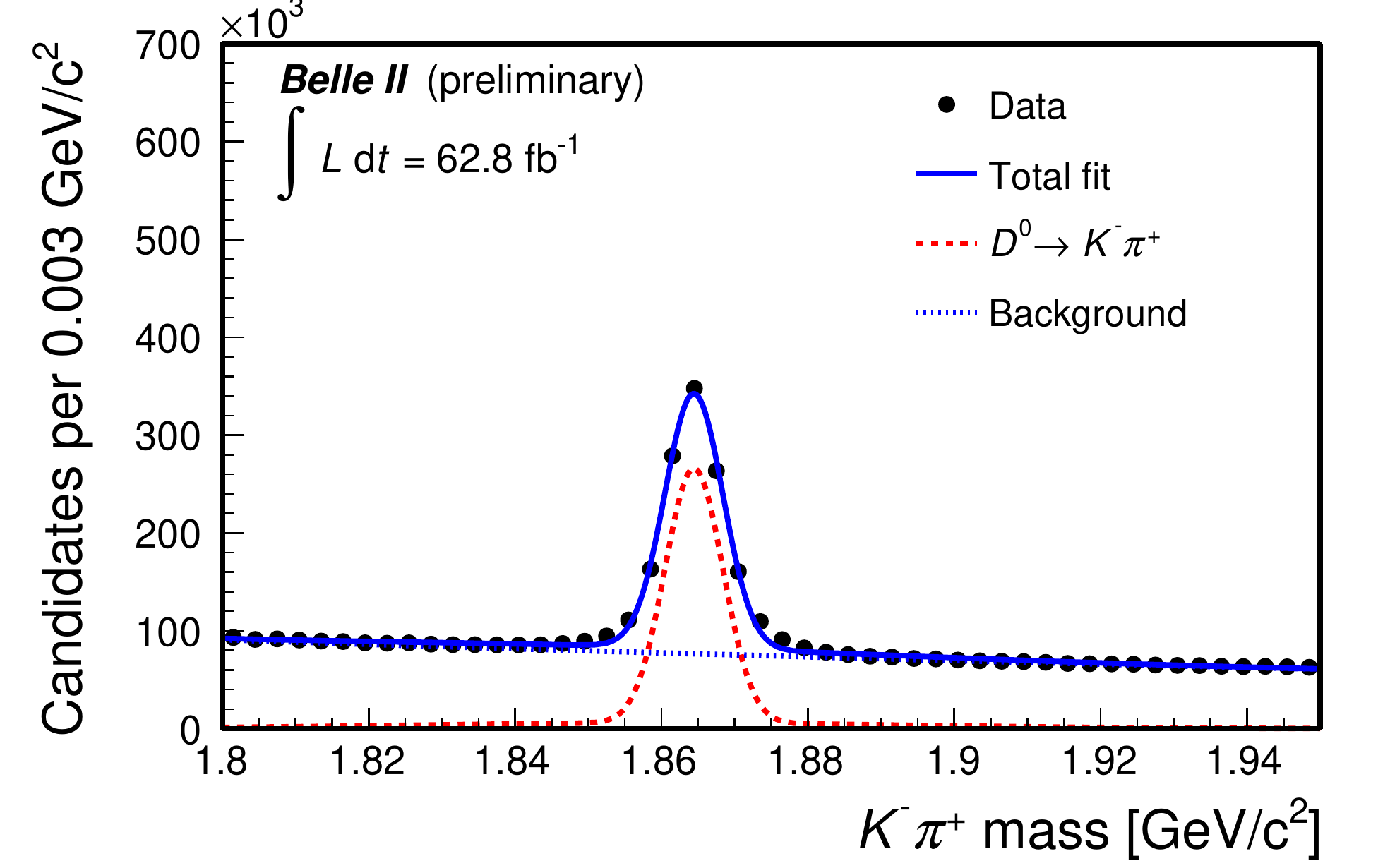}
 \includegraphics[width=0.475\textwidth]{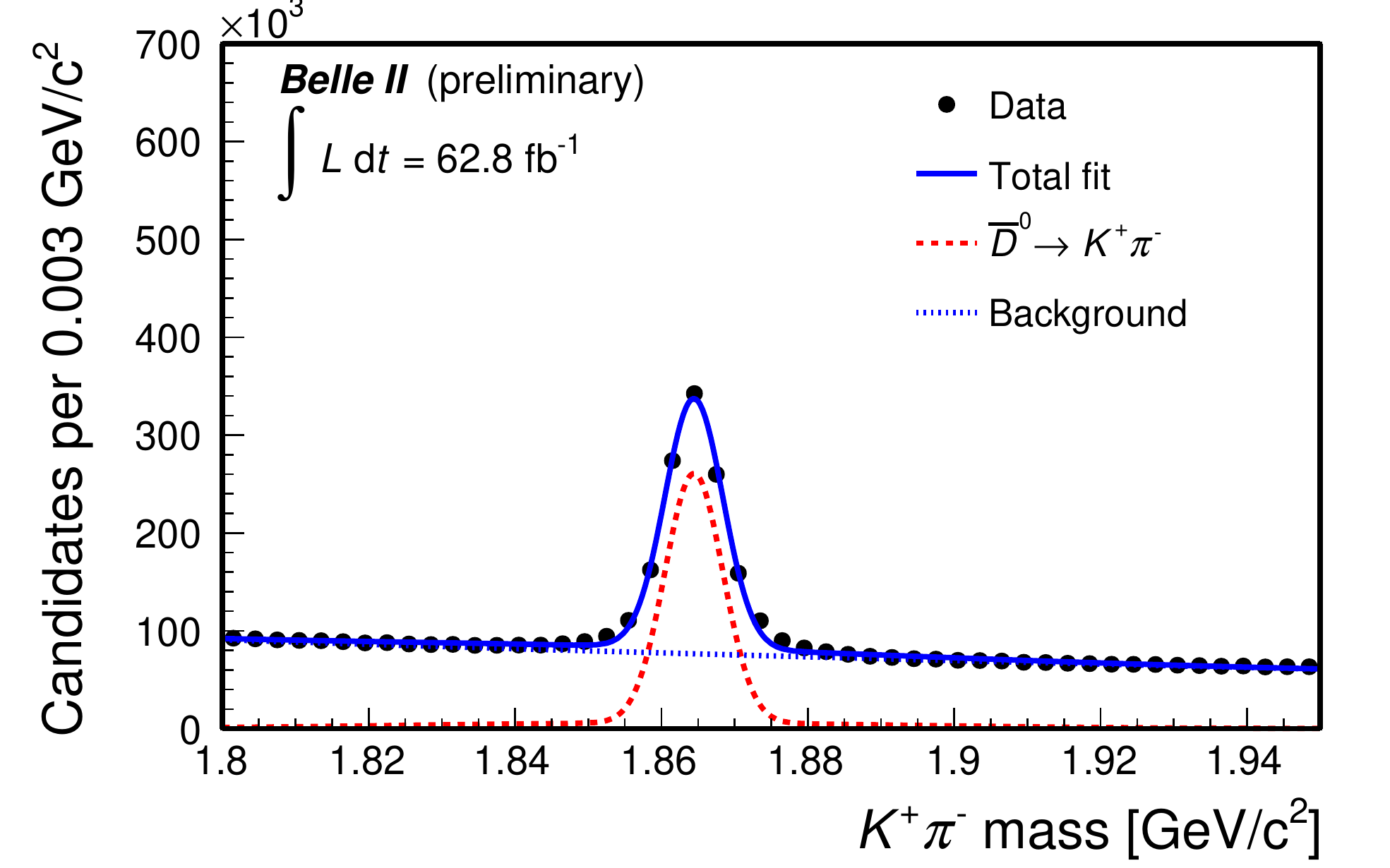}
 \caption{Distributions of $K\pi$-mass for $D^0 \to K^-\pi^+$ (left) and $\overline{D}^0 \to  K^+\pi^-$ (right) candidates reconstructed in 2019--2020 Belle~II data selected with an optimized continuum-suppression and kaon-enriching selection. Fit projections are overlaid.}
 \label{fig:DtoKpi_ACP}
\end{figure}

The raw event yields observed in data are corrected for selection and reconstruction effects to obtain physics quantities. For the measurements of branching fractions, we divide the observed yields by selection and reconstruction efficiencies. The efficiencies are determined from simulation and range between $19\%$ and $43\%$. In measurements of CP-violating asymmetries, the observed charge-specific raw event-yield asymmetries $\mathcal{A}$ are in general due to the combination of genuine CP-violating effects in the decay dynamics and instrumental asymmetries due to differences in interaction or reconstruction probabilities between opposite-charge hadrons. Such combination is additive for small asymmetries, $\mathcal{A}=\mathcal{A}_{\rm CP}+\mathcal{A}_{\rm det}$, with
\begin{equation*}
    \mathcal{A}_{\rm det}(X)=\frac{X-\bar{X}}{X+\bar{X}},
\end{equation*}
where $X$ corresponds to a given final  state and $\overline{X}$ to its charge-conjugate.
Hence, observed raw charge-specific decay yields need be corrected for instrumental effects to determine the genuine CP-violating asymmetries. We estimate the instrumental asymmetry associated with the reconstruction of $K^\pm\pi^\mp$ pairs by measuring the charge-asymmetry in an abundant sample of $D^0 \to K^- \pi^+$ decays. For these decays, direct CP-violation is expected to be smaller than 0.1\%~\cite{PDG}. We therefore attribute any nonzero asymmetry to instrumental charge asymmetries. Figure~\ref{fig:DtoKpi_ACP} shows the $K^\pm\pi^\mp$-mass  distributions for $D^0 \to K^-\pi^+$ and $\overline{D}^0 \to  K^+\pi^-$ candidates with fit projections overlaid. The resulting $K^\pm\pi^\mp$ asymmetry is directly applied to the raw measurements of charge-dependent decay rates in $B^0 \to K^+\pi^-$ to extract the physics asymmetry.

\begin{figure}[htb]
 \centering
 \includegraphics[width=0.475\textwidth]{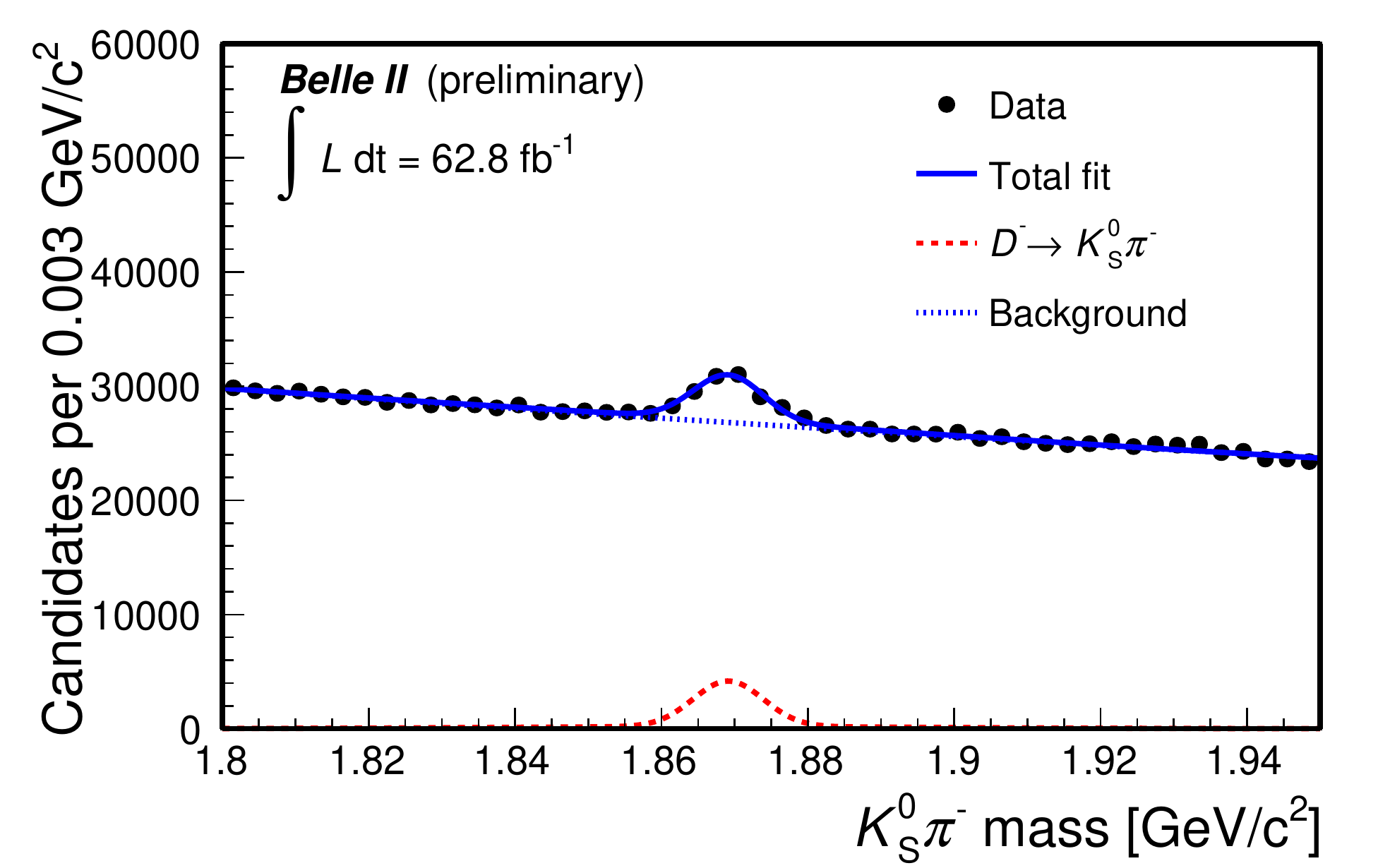}
 \includegraphics[width=0.475\textwidth]{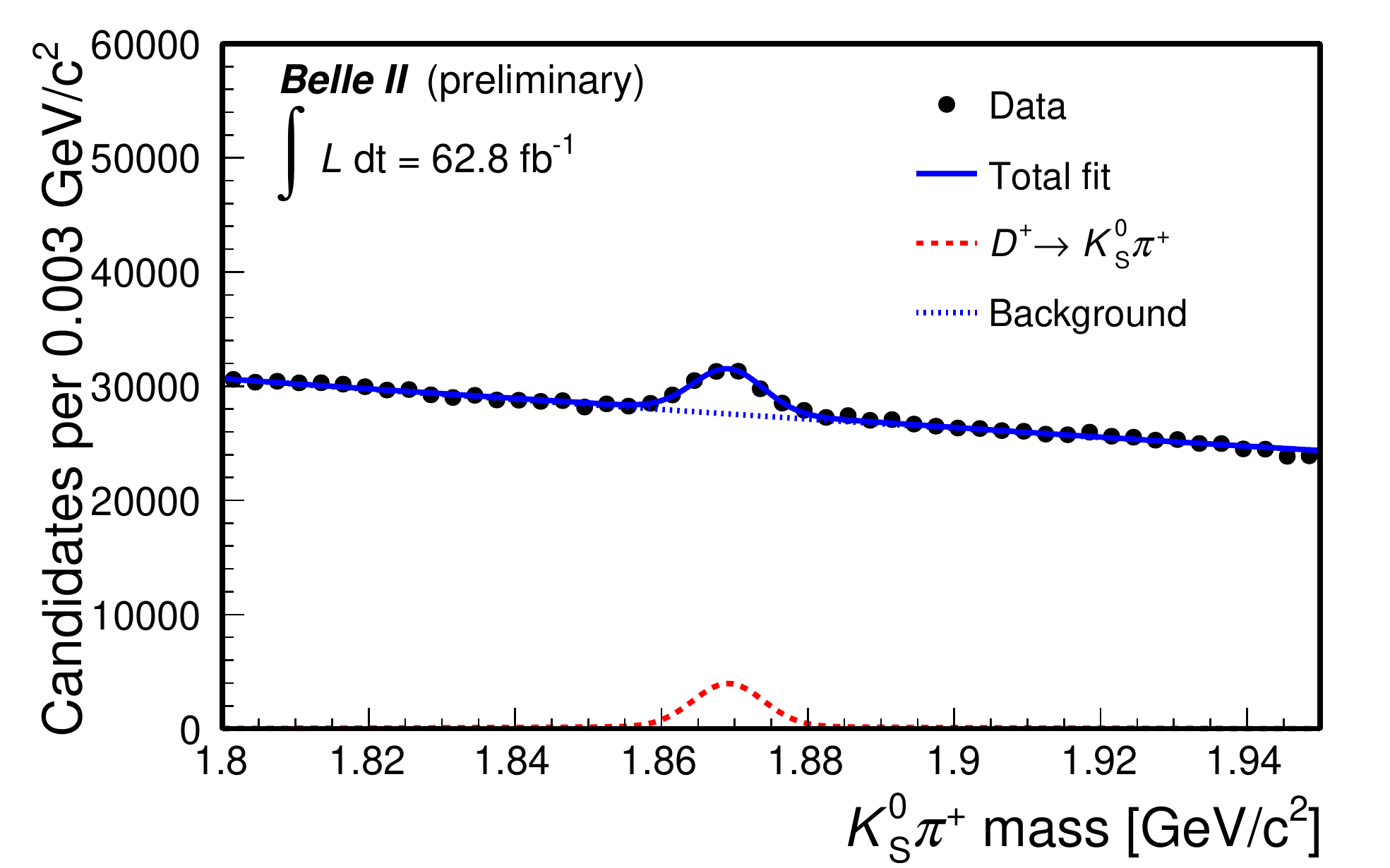}
 \caption{Distributions of $\PKzS\pi$-mass for (left) $D^- \to \PKzS\pi^-$ and (right) $D^+ \to  \PKzS\pi^+$ candidates reconstructed in 2019--2020 Belle~II data selected with an optimized continuum-suppression selection. Fit projections are overlaid.}
 \label{fig:DtoKspi_ACP}
\end{figure}

We correct the observed raw yield asymmetry of $B^+ \to \PKzS \pi^+$ decays using the yield asymmetry observed in an abundant sample of $D^+ \to \PKzS \pi^+$ decays~(Fig.~\ref{fig:DtoKspi_ACP}), in which direct CP-violation in $D^+ \to \PKzS \pi^+$ decays is expected to vanish. In each case, control channel selections are tuned to reproduce the kinematic conditions of the charmless final states that receive the corrections. Table~\ref{tab:InstrChargeAsym} shows the resulting corrections.
\begin{table}[htb]
\begin{tabular}{l  c  c}
\hline\hline
Instrumental asymmetry & Value \\
\hline
$\mathcal{A_{\rm det}}(K^+\pi^-)$   & $-0.010 \pm 0.001$ \\
$\mathcal{A_{\rm det}}(\PKzS\pi^+)$   & $+0.026 \pm 0.019$ \\
\hline\hline
\end{tabular}
\caption{Instrumental charge-asymmetries associated with $K^\pm\pi^\mp$ and $\PKzS\pi^\pm$ reconstruction, obtained using samples of $D^0 \to K^- \pi^+$ and $D^+ \to \PKzS \pi^+$ decays.}
\label{tab:InstrChargeAsym}
\end{table}

\clearpage

\section{Determination of branching fractions and CP-asymmetries}
We determine each branching fraction as 
\begin{equation*}
    \mathcal{B} = \frac{N}{\varepsilon\times 2\times N_{\PB\APB}},
\end{equation*}
where $N$ is the signal yield obtained from the fits; $\varepsilon$ is the reconstruction and selection efficiency; and $N_{\PB\APB}$ is the number of produced ${\PB\APB}$~pairs, corresponding to $35.8$~million for $\PBplus\PBminus$ and $33.9$~million for $\PBzero\APBzero$ pairs. We obtain the number of ${\PB\APB}$~pairs from the measured integrated luminosity, the \mbox{$\Pep\Pem\to\Upsilon(4{\rm S})$} cross section $(1.110 \pm 0.008)$ nb~\cite{Bevan:2014iga}, assuming that the $\Upsilon(4{\rm S})$ decays exclusively to ${\PB\APB}$~pairs, and the \mbox{$\Upsilon(4{\rm S})\to\PBzero\APBzero$}~branching fraction \mbox{$f^{00} = 0.487\pm 0.010\pm 0.008$}~\cite{Aubert:2005bq}. For the $\PBplus \to K^{0} \pi^+$ branching fraction measurement, we use the precisely determined value of 69.20\% for $\mathcal{B}(\PKzS \to \pi^+\pi^-)$~\cite{PDG} and include a factor of $1/2$ to account for the $K^{0} \to \PKzS$ transition probability.

The determination of CP-violating asymmetries is more straightforward because all factors that impact symmetrically bottom and anti-bottom rates cancel, and only flavor-specific yields and flavor-specific efficiency corrections are relevant. Table~\ref{tab:EffYieldBFSummary} shows the resulting branching fraction and direct CP-violation parameter measurements.

\begin{table}[!ht]
    \centering
\begin{tabular}{l   r  r   r  r}
\hline\hline
Decay & \multicolumn{1}{c}{$\varepsilon\times\mathcal{B}_{\rm s} [\%]$} & \multicolumn{1}{c}{Yield} & \multicolumn{1}{c}{$\mathcal{B}\,[10^{-6}]$} & \multicolumn{1}{c}{$\mathcal{A_{\rm CP}}$} \\\hline
  $B^0 \to K^+\pi^-$    & $43.0\quad$ & $568^{+29}_{-28}\;\;$  &	$17.9^{+0.9}_{-0.9}\;$ & $-0.16 \pm 0.05$\\ 
  $B^0 \to \pi^+\pi^-$  & $29.4\quad$ & $115^{+14}_{-13}\;\;$   &	$5.8 ^{+0.7}_{-0.7}\;$ & --\\ 
  $B^+ \to K^0 \pi^+$   & $6.7\quad$	& $103^{+11}_{-10}\;\;$    &  $21.4 ^{+2.3}_{-2.2}\;$ & $-0.01\pm 0.08$\\ 
   \hline\hline
\end{tabular}
    \caption{Products of selection efficiencies ($\varepsilon$) and sub-decay branching fractions ($\mathcal{B}_{\rm s}$), signal yields in 2019-2020 Belle~II data, resulting branching fractions and corrected charge-asymmetries. Only the statistical contributions to the uncertainties are shown.} 
    \label{tab:EffYieldBFSummary}
\end{table}{}

\section{Systematic uncertainties}
We consider several sources of systematic uncertainties. We assume the sources to be independent and add in quadrature the corresponding uncertainties. An overview of the effects considered follows. A summary of the fractional size of systematic uncertainties is given in tables~\ref{tab:SystematicsBF_overview} and \ref{tab:SystematicsAcp_overview}.

\subsection{Tracking efficiency} 
We assess a systematic uncertainty associated with possible data-simulation discrepancies in the reconstruction of charged particles~\cite{Bertacchi:2020eez}. The tracking efficiency in data agrees with the value observed in simulation within a $0.91\%$ uncertainty, which we linearly add as systematic uncertainty for each final-state charged particle.

\subsection{\PKzS~reconstruction efficiency} 
A small decrease, approximately linear with flight length, in $\PKzS$ reconstruction efficiency was observed in early Belle~II data with respect to simulation.
We assess a systematic uncertainty based on dedicated studies performed for the $B \to \phi K^{(*)}$~analysis~\cite{AlePhiK:2020}. We apply an uncertainty of $0.31\%$ for each centimeter of average flight length of the $\PKzS$~candidate, resulting in a 4\% total systematic uncertainty, approximately.

\subsection{Particle-identification and continuum-suppression efficiencies} We evaluate possible data-simulation discrepancies in the particle identification and in the continuum-suppression distributions using the control channel \mbox{$\PBplus\to\APDzero(\to\PKp\Pgpm)\,\Pgpp$}. The selection efficiencies obtained in data and simulation agree within $0.7\%-0.8\%$ uncertainties (depending on the selection), which are taken as systematic uncertainties.

\subsection{Number of $\PB\APB$~pairs} We  assign  a  $1.4\%$  systematic  uncertainty on the number of $\PB\APB$~pairs,  which  includes  the uncertainty  on  cross-section,  integrated  luminosity~\cite{Abudinen:2020}, and  potential  shifts  from  the peak center-of-mass energy during the run periods.

\subsection{Signal and peaking background modeling} 
Because we used empirical fit models for signal, we assess a systematic uncertainty associated with the model choice. We use ensembles of simplified simulated experiments, where the distribution for signal and background models are generated according to the default fitting model or to alternative models. We fit the composition of all simulated samples using the same likelihood as for the data and use the difference between the means of the resulting signal-yield estimates to determine a systematic uncertainty of $4.2\% - 4.9\%$ for the branching-fraction measurement. For the charge-specific fit, we fit using an alternative model and use the difference in the determined asymmetry between the alternative and nominal fit to determine the systematic uncertainty.

\subsection{Continuum background modeling}
We apply the same procedure to assess the effect of possible continuum background mismodeling, obtaining uncertainties of typically $0.3\% - 0.6\%$ for the branching-fraction measurement. For the charge-specific fit, we allow the background shapes to float independently for each charge case and assign the difference in charge-asymmetry as a systematic.

\subsection{\BB background ratio}
To assess a systematic uncertainty stemming from possible MC mismodeling of the \BB background contribution, we use ensembles of simplified simulated experiments, where the normalization of the \BB background is varied by $\pm 50\%$. We fit the composition of all simulated samples using the same likelihood as for the data and use the difference between the means of the resulting signal-yield estimates to determine a systematic uncertainty of $1.4\% - 2.1\%$ for the branching-fraction measurement. For the charge-asymmetry measurement, we vary the assumed asymmetry in the fit to data by $\pm50$\% and assign the difference in charge-asymmetry as a systematic.

\subsection{Instrumental asymmetries}
We consider the uncertainty on the values of $\mathcal{A}_{\rm det}$~(Table~\ref{tab:InstrChargeAsym}) as systematic uncertainty due to instrumental asymmetry corrections in measurements of CP~asymmetries.

\begin{table}[h]
\centering
\footnotesize
\begin{tabular}{l c c c}
\hline\hline
 Source & $\PKp\Pgpm$ & $\PKzS\Pgpp$ & $\Pgpp\Pgpm$ \\
\hline
Tracking                        & 1.82\%    & 2.73\%    & 1.82\% \\
$\PKzS$ efficiency              & -     & 4.48\%   & - \\
PID and CS eff.                 & 0.8\%    & 0.9\%  & 0.7\% \\
$N_{B\bar{B}}$                  & 1.4\%   & 1.4\%    & 1.4\%\\
Signal \& peak. bkg. model      & 4.2\%    & 4.9\%    & 4.4\%\\
Continuum bkg. model            & 0.3\%    & 0.3\%    & 0.6\% \\
$\PB\APB$ bkg. model            & 1.4\%   & 2.21\%      & - \\
\hline
Total                           & 5.1\%    & 7.7\%    & 5.0\% \\
\hline\hline
\end{tabular} 
\caption{Relative systematic uncertainties in the branching fraction measurements using the full 62.8~fb$^{-1}$ data set.}
\label{tab:SystematicsBF_overview}
\end{table}

\begin{table}[h]
\centering
\footnotesize

\begin{tabular}{l c c c c c c}
\hline\hline
 Source & $\PKp\Pgpm$ & $\PKzS\Pgpp$ \\
\hline
 Signal model   & $<$0.001 & $<$0.001\\
 Continuum background model & $<$0.001 & 0.027\\
 $\PB\APB$ background asymmetry & 0.012 & 0.043\\
 Instrumental asymmetry corrections & 0.001 & 0.019\\
\hline
 Total  & 0.012  & 0.054\\
\hline\hline
\end{tabular} 
\caption{Systematic uncertainties in the $\mathcal{A_{\rm CP}}$ measurements using the full 62.8~fb$^{-1}$ data set.}
\label{tab:SystematicsAcp_overview}
\end{table}

\section{Summary}
We report updated measurements of branching fractions and CP-violating charge asymmetries in charmless $B$ decays at Belle II. We use samples of 2019 and 2020 data corresponding to $62.8\,\si{fb^{-1}}$ of integrated luminosity. The samples are analysed using two-dimensional fits in $\Delta E$ and $M_{\it bc}$ to determine signal yields of approximately 568, 103, and 115~decays for the channels \mbox{$B^0 \to K^+\pi^-$}, \mbox{$B^+ \to \PKzS\pi^+$}, and \mbox{$B^0 \to \pi^+\pi^-$}, respectively. The signal yields are corrected for efficiencies determined from simulation and a control data sample to obtain the following branching fraction results,
\begin{center}
$\mathcal{B}(B^0 \to K^+\pi^-) = [18.0 \pm 0.9(\rm stat) \pm 0.9(\rm syst)]\times 10^{-6}$,
\end{center}

\begin{center}
$\mathcal{B}(B^+ \to K^0\pi^+) = [21.4 ^{+2.3}_{-2.2}(\rm stat) \pm 1.6(\rm syst)]\times 10^{-6}$,
\end{center}

\begin{center}
$\mathcal{B}(B^0 \to \pi^+\pi^-) = [5.8 \pm 0.7(\rm stat) \pm 0.3(\rm syst)]\times 10^{-6}$,
\end{center}

and CP-violating rate asymmetries,

\begin{center}
$\mathcal{A_{\rm CP}}(B^0 \to K^+\pi^-) = -0.16 \pm 0.05(\rm stat) \pm 0.01(\rm syst)$, \\
\end{center}
\begin{center}
$\mathcal{A_{\rm CP}}(B^+ \to \PKzS\pi^+) = -0.01 \pm 0.08(\rm stat) \pm 0.05(\rm syst)$.
\end{center}
The results are compatible with known determinations and contribute important information to an early assessment of Belle II detector performance.

\clearpage
\section*{Acknowledgments}
We thank the SuperKEKB group for the excellent operation of the
accelerator; the KEK cryogenics group for the efficient
operation of the solenoid; the KEK computer group for
on-site computing support; and the raw-data centers at
BNL, DESY, GridKa, IN2P3, and INFN for off-site computing support.
This work was supported by the following funding sources:
Science Committee of the Republic of Armenia Grant No. 20TTCG-1C010;
Australian Research Council and research grant Nos.
DP180102629, 
DP170102389, 
DP170102204, 
DP150103061, 
FT130100303, 
FT130100018,
and
FT120100745;
Austrian Federal Ministry of Education, Science and Research,
Austrian Science Fund No. P 31361-N36, and
Horizon 2020 ERC Starting Grant no. 947006 ``InterLeptons''; 
Natural Sciences and Engineering Research Council of Canada, Compute Canada and CANARIE;
Chinese Academy of Sciences and research grant No. QYZDJ-SSW-SLH011,
National Natural Science Foundation of China and research grant Nos.
11521505,
11575017,
11675166,
11761141009,
11705209,
and
11975076,
LiaoNing Revitalization Talents Program under contract No. XLYC1807135,
Shanghai Municipal Science and Technology Committee under contract No. 19ZR1403000,
Shanghai Pujiang Program under Grant No. 18PJ1401000,
and the CAS Center for Excellence in Particle Physics (CCEPP);
the Ministry of Education, Youth and Sports of the Czech Republic under Contract No.~LTT17020 and 
Charles University grants SVV 260448 and GAUK 404316;
European Research Council, 7th Framework PIEF-GA-2013-622527, 
Horizon 2020 ERC-Advanced Grants No. 267104 and 884719,
Horizon 2020 ERC-Consolidator Grant No. 819127,
Horizon 2020 Marie Sklodowska-Curie grant agreement No. 700525 `NIOBE,' 
and
Horizon 2020 Marie Sklodowska-Curie RISE project JENNIFER2 grant agreement No. 822070 (European grants);
L'Institut National de Physique Nucl\'{e}aire et de Physique des Particules (IN2P3) du CNRS (France);
BMBF, DFG, HGF, MPG, and AvH Foundation (Germany);
Department of Atomic Energy under Project Identification No. RTI 4002 and Department of Science and Technology (India);
Israel Science Foundation grant No. 2476/17,
United States-Israel Binational Science Foundation grant No. 2016113, and
Israel Ministry of Science grant No. 3-16543;
Istituto Nazionale di Fisica Nucleare and the research grants BELLE2;
Japan Society for the Promotion of Science,  Grant-in-Aid for Scientific Research grant Nos.
16H03968, 
16H03993, 
16H06492,
16K05323, 
17H01133, 
17H05405, 
18K03621, 
18H03710, 
18H05226,
19H00682, 
26220706,
and
26400255,
the National Institute of Informatics, and Science Information NETwork 5 (SINET5), 
and
the Ministry of Education, Culture, Sports, Science, and Technology (MEXT) of Japan;  
National Research Foundation (NRF) of Korea Grant Nos.
2016R1\-D1A1B\-01010135,
2016R1\-D1A1B\-02012900,
2018R1\-A2B\-3003643,
2018R1\-A6A1A\-06024970,
2018R1\-D1A1B\-07047294,
2019K1\-A3A7A\-09033840,
and
2019R1\-I1A3A\-01058933,
Radiation Science Research Institute,
Foreign Large-size Research Facility Application Supporting project,
the Global Science Experimental Data Hub Center of the Korea Institute of Science and Technology Information
and
KREONET/GLORIAD;
Universiti Malaya RU grant, Akademi Sains Malaysia and Ministry of Education Malaysia;
Frontiers of Science Program contracts
FOINS-296,
CB-221329,
CB-236394,
CB-254409,
and
CB-180023, and SEP-CINVESTAV research grant 237 (Mexico);
the Polish Ministry of Science and Higher Education and the National Science Center;
the Ministry of Science and Higher Education of the Russian Federation,
Agreement 14.W03.31.0026, and
the HSE University Basic Research Program, Moscow;
University of Tabuk research grants
S-0256-1438 and S-0280-1439 (Saudi Arabia);
Slovenian Research Agency and research grant Nos.
J1-9124
and
P1-0135; 
Agencia Estatal de Investigacion, Spain grant Nos.
FPA2014-55613-P
and
FPA2017-84445-P,
and
CIDEGENT/2018/020 of Generalitat Valenciana;
Ministry of Science and Technology and research grant Nos.
MOST106-2112-M-002-005-MY3
and
MOST107-2119-M-002-035-MY3, 
and the Ministry of Education (Taiwan);
Thailand Center of Excellence in Physics;
TUBITAK ULAKBIM (Turkey);
Ministry of Education and Science of Ukraine;
the US National Science Foundation and research grant Nos.
PHY-1807007 
and
PHY-1913789, 
and the US Department of Energy and research grant Nos.
DE-AC06-76RLO1830, 
DE-SC0007983, 
DE-SC0009824, 
DE-SC0009973, 
DE-SC0010073, 
DE-SC0010118, 
DE-SC0010504, 
DE-SC0011784, 
DE-SC0012704, 
DE-SC0021274; 
and
the Vietnam Academy of Science and Technology (VAST) under grant DL0000.05/21-23.

\clearpage
\bibliography{belle2}
\bibliographystyle{belle2-note}

\end{document}